\documentclass[traditabstract]{aa} 
\usepackage{graphicx}
\usepackage[varg]{txfonts}
\usepackage{natbib}
\def\as         {$^{\prime\prime}$}

\def\zsun       {$Z_{\odot}$}

\def\ltsima{$\; \buildrel < \over \sim \;$} 
\def\simlt{\lower.5ex\hbox{\ltsima}} 
\def\gtsima{$\; \buildrel > \over \sim \;$} 
\def\simgt{\lower.5ex\hbox{\gtsima}} 
\begin{document}
\title{Metal distribution in sloshing galaxy clusters: the case of A496.}
\subtitle{}
\author{Simona Ghizzardi 
          \inst{1}
          \and
          Sabrina De Grandi \inst{2} \and Silvano Molendi \inst{1}          }
\institute{INAF - Istituto di Astrofisica Spaziale e Fisica Cosmica - Milano, via E. Bassini 15, I-20133 Milano, Italy\\
              \email{simona@iasf-milano.inaf.it}
         \and
             INAF - Osservatorio Astronomico di Brera, via E. Bianchi 46, I-23807 Merate (LC), Italy 
             %\email{sabrina.degrandi@brera.inaf.it}
             }
\date{Received ; accepted }
% \abstract{}{}{}{}{} 
% 5 {} token are mandatory
\abstract{
We report results from a detailed study of the sloshing gas in the core of A496. We detect
the low temperature/entropy spiral feature found in several cores, we  also find that conduction between 
the gas in the spiral and the ambient medium must be suppressed by more than one order of magnitude with respect to
Spitzer conductivity. Intriguingly, while the gas in the spiral
features a higher metal abundance than the surrounding medium, it follows the 
entropy vs metal abundance relation defined by gas lying outside the spiral. The most plausible
explanation for this behavior is that the low entropy metal rich plasma uplifted
through the cluster atmosphere by sloshing, suffers little  heating or mixing with the ambient medium.
While sloshing appears to be capable of uplifting significant amounts of gas, the limited heat exchange and 
mixing between gas in and outside the spiral implies that this mechanism is not at all effective in: 1) permanently 
redistributing metals within the core region and 2) heating up the coolest and densest gas, thereby providing little or no 
contribution to staving of catastrophic cooling in cool cores.}{}{}{}{}
%\abstract
%   {2} 
%   {3}
%   {4}
%   {5}


\maketitle
%
%________________________________________________________________

\section{Introduction}
\label{sec:intro}

Over the past decade, with the advent of the current generation of X-ray telescopes (mainly XMM-Newton and Chandra), 
high-resolution observations have revealed a wealth of small-scale substructures within the intracluster gas of 
galaxy groups and clusters. Particularly remarkable features are cold fronts \citep[see][for a detailed 
review]{MM_review:2007}. 
They appear as sharp surface brightness discontinuities accompanied by a temperature jump, where the denser 
gas is also cooler, so that the pressure is approximately continuous across the front.
Soon after their discovery in the early Chandra observations of A2142 and A3667 
\citep{Maxim:2000,Vikhlinin1:2001,Vikhlinin2:2001,Vikhlinin:2002}, 
these features have been largely observed in many clusters of galaxies. 
Indeed, cold fronts appear to be almost ubiquitous in galaxy clusters \citep[][G10 hereafter]{Maxim:2003,G10} with the 
vast majority of clusters  hosting at least one cold front and many cool-core clusters featuring more than one (G10). 
More recently, several cold fronts have been detected also in galaxy groups 
(\citealp[NGC 5044:][]{OSullivan5044:2014,Gasta5044:2009,David5044:2009,Buote:2003}; \citealp[NGC 5846:][]{Machachek5846:2011};
\citealp[3C 449 Group:][]{Lal3C449:2013};
 \citealp[IC1860:][]{GastaIC1860:2013}; \citealp[Pegasus Group:][]{Randall2pegasus:2009}; \citealp[NGC5098:][]{Randall5098:2009}).
Since understanding the nature of such a widespread phenomenon is mandatory for characterizing the dynamics of galaxy clusters, 
cold fronts have been largely investigated 
in the last years by several authors, using X-ray observations 
\citep[e.g.][among the most recent works]{Canning:2013,PM:2013,Clarke_A2029:2013,Rossetti_A2142:2013,Ettori:2013,GastaIC1860:2013}, 
optical data \citep[e.g.][]{Owers2142.rxj:2011,Owers:2009,Owers:1202:2009} 
and numerical simulations \citep[see e.g.][]{AM06,Hallman:2010,ZuHone:2010,ZuHone:2011,Roediger_ZuHone:2012}. 

In the currently dominant picture, cold fronts in cool core clusters are believed 
to be due to the sloshing of the innermost cool gas within the underlying dark matter potential well 
\citep{TH:2005,AM06}. 
Numerical simulations show that sloshing can be triggered within cool cores by an off-axis minor merger: 
an infalling subclump induces a perturbation to the underlying gravitational potential
of the main cluster; the cooler inner gas is
displaced from the center of the potential well, lifted upwards, decoupled
from the dark matter through ram pressure, and finally starts
to slosh. The oscillations generated by the gravitational disturbance are long-lasting and can survive for Gyrs 
and produce a succession of concentric fronts, propagating outwards.
If the initial fly-by is off-center, the cool gas acquires angular momentum and the sloshing takes a spiral-like appearance.
Simulations by \citet{AM06} highlighted that a steep central entropy is a necessary condition for 
the onset of the sloshing mechanism. On the observational 
side this requirement has been confirmed by G10 \citep[see, however, the case of A2142,][]{Rossetti_A2142:2013}.

Though widely investigated, cold fronts and sloshing studies have focussed mainly on dynamics and thermodynamics, 
while the relationship between the sloshing and the cluster chemical properties has enjoyed relatively little attention. 
A detailed characterization of the metal abundance across the fronts is only available for a
handful of objects (\citealp[Perseus:][]{Fabian_Perseus:2011}; \citealp[Centaurus:][]{Sanders_Centaurus:2006}; 
\citealp[A2204:][]{Sanders_A2204:2005, Sanders_A2204:2009}; \citealp[A2052:][]{DePlaa2052:2010}; 
\citealp[A1201:][]{Ma1201:2012}; \citealp[M87:][]{SimionescuM87:2010}; \citealp[A3581:][]{Canning:2013}) and recently 
\citet{OSullivan5044:2014} find a close correlation between the metallicity distribution
and the cold fronts position in the galaxy group NGC 5044.
This is a significant limitation as sloshing may play a crucial role in redistributing metals
in the ICM. 
Notoriously, cool-core clusters have prominent metallicity peaks \citep{DGM:2001} in
their centers, consistent with being produced by the central BCG galaxy \citep[e.g.][]{DGM:2001,Degrandi:2004}. 
As a consequence, heavy elements are powerful markers of the central cool gas and
the metal distribution may trace the history of the central gas motions during the sloshing.
Several works show that the iron abundance
profile is broader than the stellar light profile of the BCG \citep[e.g.][]{DGM:2001,Graham:2006,
Rebusco:2006}, suggesting that metals are drifted away from the BGC and spread outward. 
However the mechanism responsible of the mismatch between the metal abundance and light profiles 
is still poorly understood. Sloshing may be a viable way to broaden the distribution of metal abundance profile.

In this paper we aim to characterize the sloshing mechanism through metal distribution in the ICM.
To address this issue, we choose to analyze a long (120 ksec) XMM-Newton observation of A496.
A496 is a bright, nearby $(z=0.0329)$, cool-core cluster \citep{Peres:1998}.
It is a particularly suited candidate to inspect the correlation between metal distribution and
cold fronts as it is the only cluster to host four cold fronts \citep{Ghizzardi:13p} in its circum-core area 
($\simlt 250$ kpc). In addition, A496 has no other particular feature within its core (like bubbles, rims, cavities) 
which would add complexity to the problem and to the interpretation of results. 

The paper is organized as follows: in Sec. \ref{sec:data_reduc} we
describe the XMM observation, the data reduction, the procedure used to produce the thermodynamical maps and the
spectral extraction methods used in our analysis. In Sec. \ref{sec:results} we present the detection and 
characterization of cold fronts and other sloshing signatures, including the metal abundance distribution.
We discuss results in Sec. \ref{sec:discussion} and summarize our findings in Sec. \ref{sec:summary}.
Throughout the article, if not otherwise stated, we plot and tabulate values
with errors quoted at the 68\%  ($1\sigma$) confidence level.
In our analysis, we assume a flat $\Lambda$CDM cosmology with a Hubble constant $H_0 = 70 {~\rm km} {~\rm s}^{-1} {~\rm Mpc}^{-1}$,
$\Omega_m = 0.3$, $\Omega_\Lambda = 0.7$. A496 is assumed to have a redshift $z= 0.0329$: at this distance 1 arcsec corresponds to 0.65 kpc.
All the metallicities are given relative to the
solar abundance in \citet{AG:1989}.

\section{Data sets and data processing}
\label{sec:data_reduc}

In this work we used two XMM-Newton observations of Abell 496 performed on August 11, 2007
(observation id. 0506260301) and on February 18, 2008 (0506260401) for a total nominal 
exposure time of 141 ks. 
We reprocessed the Observation Data Files (ODF) of each observation separately 
using the Science Analysis System (SAS) version 11.0.0. 
After the production of the calibrated event lists for the EPIC MOS1, MOS2 and {\it pn} data 
with {\it emchain} and {\it epchain} tasks, we performed a soft proton cleaning using a double 
filtering process.
We first removed soft protons spikes by screening the light curves produced in 100 seconds bins in the 
10--12 keV band for MOS and 10--13 keV for {\it pn} (as a safety check for possible flares 
with soft spectra) and then by applying the appropriate threshold 
for each instrument: for MOS we used slightly different thresholds in the two observations, namely  
0.20 cts s$^{-1}$ for 0506260301 and 0.15 cts s$^{-1}$  for 0506260401, for the {\it pn} 
instead we always used a threshold of 0.60  cts s$^{-1}$. These thresholds are slightly larger 
than the standard one due to a higher level of soft protons.
To exclude possible residual flares contributing below 10 keV, we extracted a light curve 
in the 2--5 keV band and fitted the histogram produced from this curve with a Gaussian distribution. 
To generate the final filtered event files we rejected all events registered at times with count rates 
larger than 3$\sigma$ from the mean of this distribution. 
We finally filtered event files according to FLAG (FLAG==0) and PATTERN (PATTERN $\le$ 12 for MOS 
and PATTERN==0 for {\it pn}) criteria. 

The resulting net exposure times for dataset 0506260301 are 57.1 ks, 58.7 ks and 28.3 ks for the 
MOS1, MOS2, and {\it pn} detectors respectively, whereas for the 0506260401 observations the exposures
are 53.0 ks, 47.8 ks and 30.0 ks for the MOS1, MOS2, and {\it pn}.

Since we are more interested in the study of the relatively high surface brightness
regions of A496, where the background subtraction is less critical than in the case of more 
external low surface brightness regions, we made use of
blank-sky fields instead of proceeding with a more detailed modeling of the different background 
components.
The blank-sky fields for EPIC MOS and {\it pn} were produced by \citet[][see Appendix B 
``The analysis of blank field observations'' in their paper]{Leccardi_temp:2008} by analyzing a large 
number of observations for a total exposure time of $\sim 700$ ks for MOS and $\sim 500$ ks for {\it pn}.
We refined our background analysis by also performing a background rescaling for each observation 
separately to account for temporal variations of the background. We estimated the background 
intensity from spectra extracted from an external ring between 10$^\prime$ and 12$^\prime$ 
centered on the emission peak at (RA, DEC) = (4:33:38; -13:15:41),
taking into account only the 10--12 keV band (to avoid possible extended cluster emission residuals 
in this region). 
Finally, we rescaled the blank-sky fields background to the local value. The scaling factors 
are in the range [1.5-1.8]. These values are larger than usual because of the larger threshold adopted
for filtering soft protons. 
This procedure scales both the instrumental and the sky background. However, the sky background in 
the central regions of the cluster, where our analysis is focussed, is almost negligible, so this procedure 
does not introduce any significant bias in the measure.

\subsection{X-ray surface brightness image and thermodynamical maps}
\label{sec:xmaps}
\noindent

\begin{figure}
\centering
\includegraphics[angle=270, width=9 truecm]{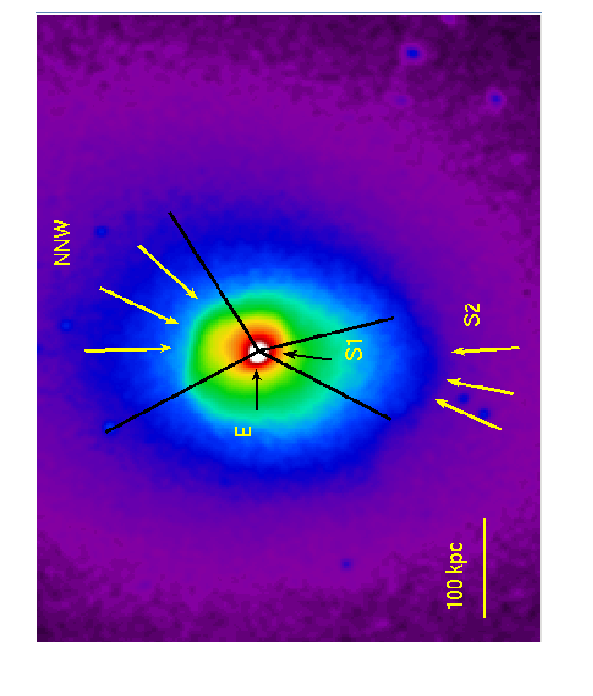}
\caption{EPIC surface brightness image in the energy range 0.4 -2 keV. Black lines define the sectors hosting 
cold fronts (see text for further details).
Black arrows mark the position of the inner cold fronts (labeled E and S1) and 
yellow arrows mark the positions of the outermost cold fronts (labeled NNW ans S2).}
\label{fig:mappa_epic_fx}
\end{figure}

Using the cleaned event files for MOS1, MOS2, and {\it pn} we built the 
EPIC flux map: MOS1 + MOS2 + {\it pn} (with {\it pn} images corrected for out of time events). 
This flux image is computed in the $0.4 - 2$ keV energy band and the two observations were
joined to obtain a single mosaicized map. Details on the preparation of the EPIC flux images can be 
found in 
G10 and \citet{Rossetti_A3558:2007}.
The resulting flux map is shown in Fig. \ref{fig:mappa_epic_fx}. 
The flux image is in units of $ 10^{-15} {\rm erg ~ cm}^{-2} {\rm s}^{-1} {\rm pixel}^{-1}$ (one pixel 
is
$3.25 \times 3.25$ arcsec$^2$). 

\begin{figure*}
  %\centering
   \includegraphics[angle=0,width=18.3 truecm]{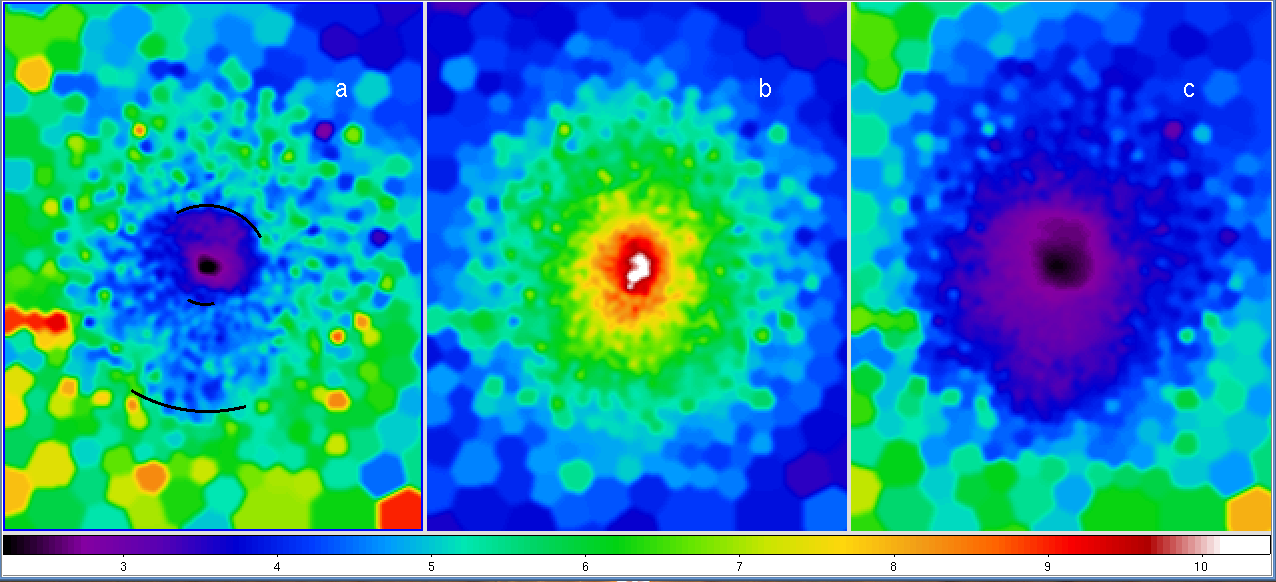}
   \caption{EPIC temperature (a), projected pressure (b) and projected entropy (c) maps for A496. Maps have been binned 
using a WVT+BBF algorithm with a S/N=10 per bin and then smoothed with a $\sigma = 9.75^{\prime\prime}$ Gaussian. 
Black arcs in panel (a) mark the cold fronts positions.
The color bar indicates the temperature in keV.}
              
\label{fig:WVT_mappe_10}%
    
\end{figure*}

To study the discontinuities in A496, we also built thermodynamical maps using 
the WVT + Broad Band Fitting method introduced 
and described in \citet{Rossetti_A3558:2007}, where we use the Weighted Voronoi Tessellation (WVT) binning 
algorithm by \citet{Diehl:2006}, in place of
the \citet{Cappellari:2003} Voronoi binning algorithm.
We set a minimum Signal/Noise value $S/N=10$ for the binning algorithm and we used the energy 
bands ($0.4 - 0.8$ keV,
$0.8 - 1.2$ keV, $1.2 - 2$ keV, $2 - 4$ keV, $4 - 10$ keV) for the broad band fitting procedure.
In Fig. \ref{fig:WVT_mappe_10} we show the maps for the temperature $T$ (panel a), 
the projected pressure $P$ (panel b) and the projected entropy $K$ (panel c).

As is conventional in X-ray
astronomy, we quantify the entropy using the adiabatic constant
$K = kTn_e^{-2/3}$
($T$ and $n_e$ are the gas temperature and density respectively
and $k$ is the Boltzmann constant) following \citet[see also \citealp{Ponman_entropy_sc:2003}]{Voit_entropy:2005}.
The specific entropy $s$ is related to $K$ through the relation
$s \propto lnK$. For brevity, we will refer to $K$ as entropy throughout the paper.
To derive $K$, we should deproject cluster surface brightness and temperature (see e.g. 
\citealp{Ghizzardi_M87:2004}).
However, because of the presence of surface brightness discontinuities, the cluster is asymmetric
and the deprojection technique is not particularly useful. For this reason, we will use the projected entropy
derived as $T/EM^{1/3} $ and the projected pressure derived as $T\cdot EM^{1/2}$ 
where $T$ and $EM$ are the temperature and the emission measure (XSPEC normalization per pixel) derived from the
spectral fit.  Since we use projected quantities, $P$ and $K$ are given in arbitrary units throughout the paper. 

\subsection{Spatially resolved spectral analysis}
\label{sec:spec_analysis}
\noindent

We performed the spectral analysis of A496 in a series of regions, namely radial bins along sectors
centered on the emission peak (see Sect. \ref{sec:t_z_prof}) and other relevant polygonal regions 
(see Sect. \ref{sec:spiral}). 
In the following we describe how we handled the spectrum of a generic region regardless of its shape.

We extracted the total (i.e., source plus background) spectrum of each region for each observation.
The corresponding redistribution matrix files (RMFs) and ancillary response files (ARFs) were generated 
with the SAS tasks  {\it rmfgen} 
and {\it arfgen}.
The background spectrum was extracted from blank-sky fields as described at the beginning of this 
Section.
Before proceeding with the spectral analysis,
we co-added, by weighting appropriately, the total and background spectra of the two observations 
with the FTOOLS {\it matpha}, the ARFs with {\it addarf} and the RMFs with {\it addrmf}. 
Spectra from all three EPIC instruments were then fitted simultaneously
using the spectral band between 0.7 and 8.0 keV.
All spectral fits were performed with the XSPEC package (version 11.3.2, \citealt{Arnaud_XSPEC:1996}).

We analyzed each cluster spectrum with two different models: 
a one temperature thermal model with the plasma in collisional ionization equilibrium 
({\it vapec} model in XSPEC), referred to as the 1T model hereafter, 
and, a multi-temperature model in which the temperature
distributions are a Gaussian (GDEM model,
as described in \citealp{Buote:2003} eq. 3), where the mean temperature $T_0$ 
is the mean and $\sigma_T$ is the standard deviation of the Gaussian.

In the 1T model we allowed temperature, normalization, redshift and 
elements Si, S and Fe to vary freely. C, N, O were fixed at 0.4;  Mg and Al were linked to Ne 
which is left free to vary; Ar and Ca were linked to S 
as these elements are less abundant and we do not plan to study them in this work. 
Ni was fixed at 1.8 times the value of Fe 
following the result found by \citet{DGM:2009} from a sample of cool core clusters.
In the GDEM model we left as free parameters the mean temperature T$_0$ and the
 $\sigma_T$ of the Gaussian temperature distribution and the metallicity. 
In both models the redshift was always left free, to account for small calibration differences between 
the two MOS and the {\it pn} cameras. We accounted for low energy absorption by 
multiplying by the Galactic hydrogen column density, 
$N_H = 3.76\times 10^{+20}$ cm$^{-2}$, 
determined by HI surveys \citep{Kalberla:2005} through the {\it wabs} absorption model in XSPEC. 
In the following Sections we report results from spectral fits with $N_H$ fixed at the weighted Galactic 
value only. We have also allowed $N_H$ to vary in both models finding no significant differences in the 
derived temperature and metal abundances values.

As pointed out by \citet{Leccardi_temp:2008}, when fitting spectra with XSPEC it is appropriate 
to allow the metallicities to assume negative values. This procedure is necessary to avoid 
underestimating metal abundances, especially 
in the case of low metallicity, statistically poor spectra (for a more detailed discussion of this point 
see Appendix A in \citealt{Leccardi_temp:2008}).
Abundances are measured relative to the solar photospheric values of \citet{AG:1989}, 
where Fe $= 4.68\times 10^{-5}$, Si $= 3.55\times 10^{-5}$ and S $= 1.62\times 10^{-5}$ 
(by number relative to H). We have chosen these values to allow direct comparison with other works 
present in the literature. 

Finally, in the case of the {\it pn} only, we always include a multiplicative component that performs a 
Gaussian smoothing of the spectral model ({\it gsmooth} in XSPEC) to take into account a small 
mis-calibration of the redistribution matrix of this detector as shown first in \citet{MG:2009} 
and then investigated in details by \citet[see their Sect. 4.1 and Fig. 1]{DGM:2009}.

\section{Results}
\label{sec:results}

\subsection{4 circum-core cold fronts in A496}
\label{sec:4cf}

A visual inspection of the flux map of Fig. \ref{fig:mappa_epic_fx} shows a very regular, 
centrally peaked X--ray surface brightness typical of a cool core cluster.
The surface brightness is fairly elongated along the N-NW to S-SE direction.
The map clearly reveals several sharp discontinuities.
The main discontinuity is located in the NNW direction (30$^{\circ}$-120$^{\circ}$; angles are measured from W) at 
a distance $\sim$ 65 kpc 
($\sim$ 100\as) from the X--ray peak. The front is sharp with the surface brightness dropping by about 
a factor of 3 in 20 kpc and exhibits a boxy morphology.
A surface brightness discontinuity is also observed $\sim$ 35 kpc ($\sim$ 55\as) from the center on the opposite 
side, in the south direction (labeled S1 in Fig. \ref{fig:mappa_epic_fx}).
Both these discontinuities had also been detected and studied by \citet{Dupke:2003} and \citet{DW:2007} 
using Chandra data.
The former authors also detected a third discontinuity located at
$\sim$16 kpc (labeled E in Fig. \ref{fig:mappa_epic_fx}) east of the center.
This cold front is too close to the center to be resolved by the XMM-Newton instruments.

Another outer discontinuity, marked S2 in Fig. \ref{fig:mappa_epic_fx},
is detected further south, (240$^{\circ}$-285$^{\circ}$), $\sim$ 160 kpc ($\sim$ 240\as) from the 
center (see also \citealp{Tanaka:2006}).
This front could not be detected in Chandra maps as it lies almost at the edge of the ACIS-S3 chip.

In Fig. \ref{fig:WVT_mappe_10} we show the maps for the temperature (panel a), 
the projected pressure (panel b) and the projected entropy (panel c). 
The black arcs in panel (a) mark the position of the discontinuities.

The temperature map confirms that the cluster has a central cool core with the temperature decreasing 
in the central regions, 
reaching the value of $\sim 2$ keV in the core, starting from an outer value of $\sim 5$ keV.
The map also shows that the cooler gas lies in the higher density side of the surface brightness 
edges with the temperature increasing across the discontinuity as typical of cold fronts features.

\begin{figure}
 %  \centering
{{ %\includegraphics[width=0.45\textwidth]{figs/SB_prof_30_75.ps}}
\includegraphics[width=0.25\textwidth]{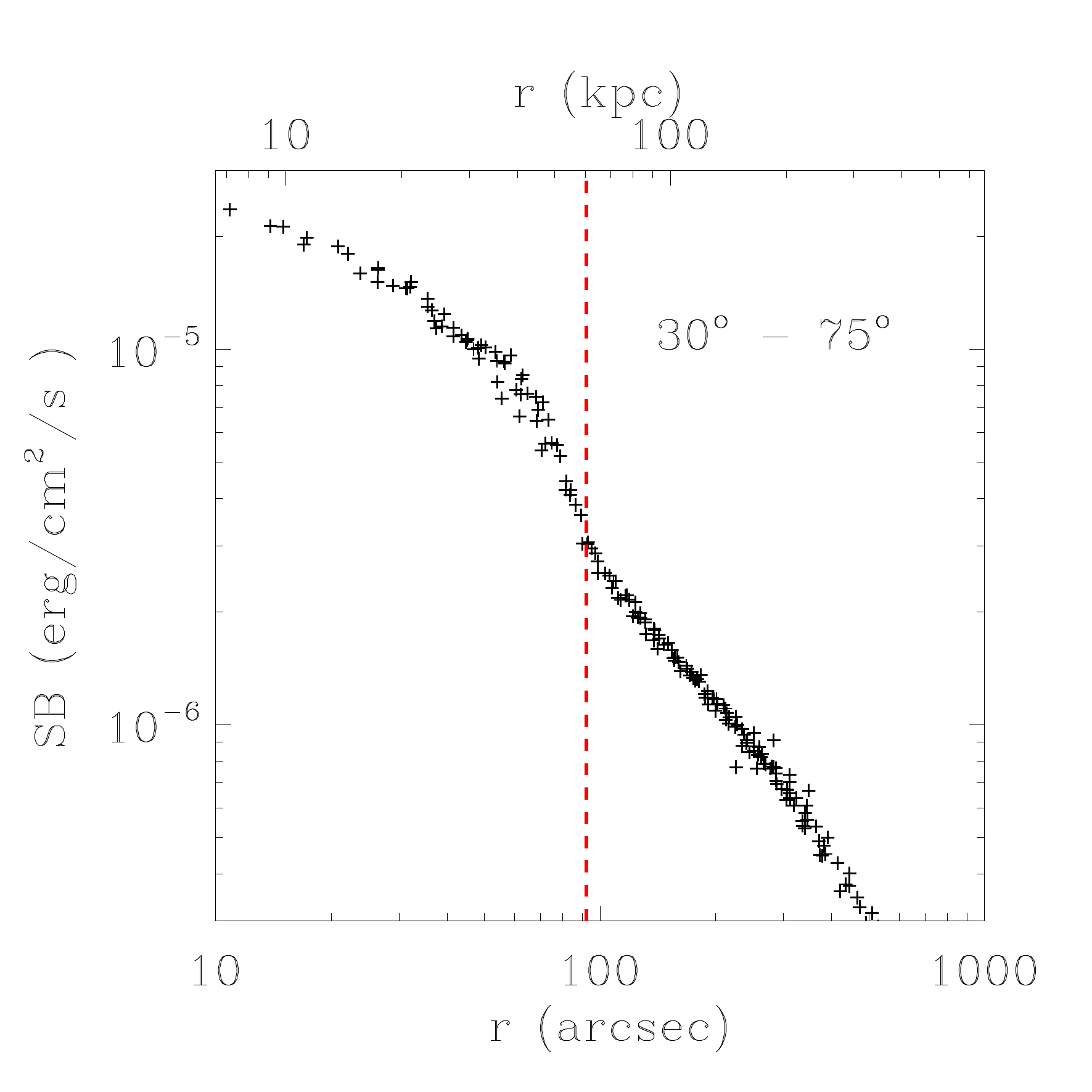}} %0.45 in referee %0.25 regular
%%\hspace 5mmm
\hspace{-7truemm}
{\includegraphics[width=0.25\textwidth]{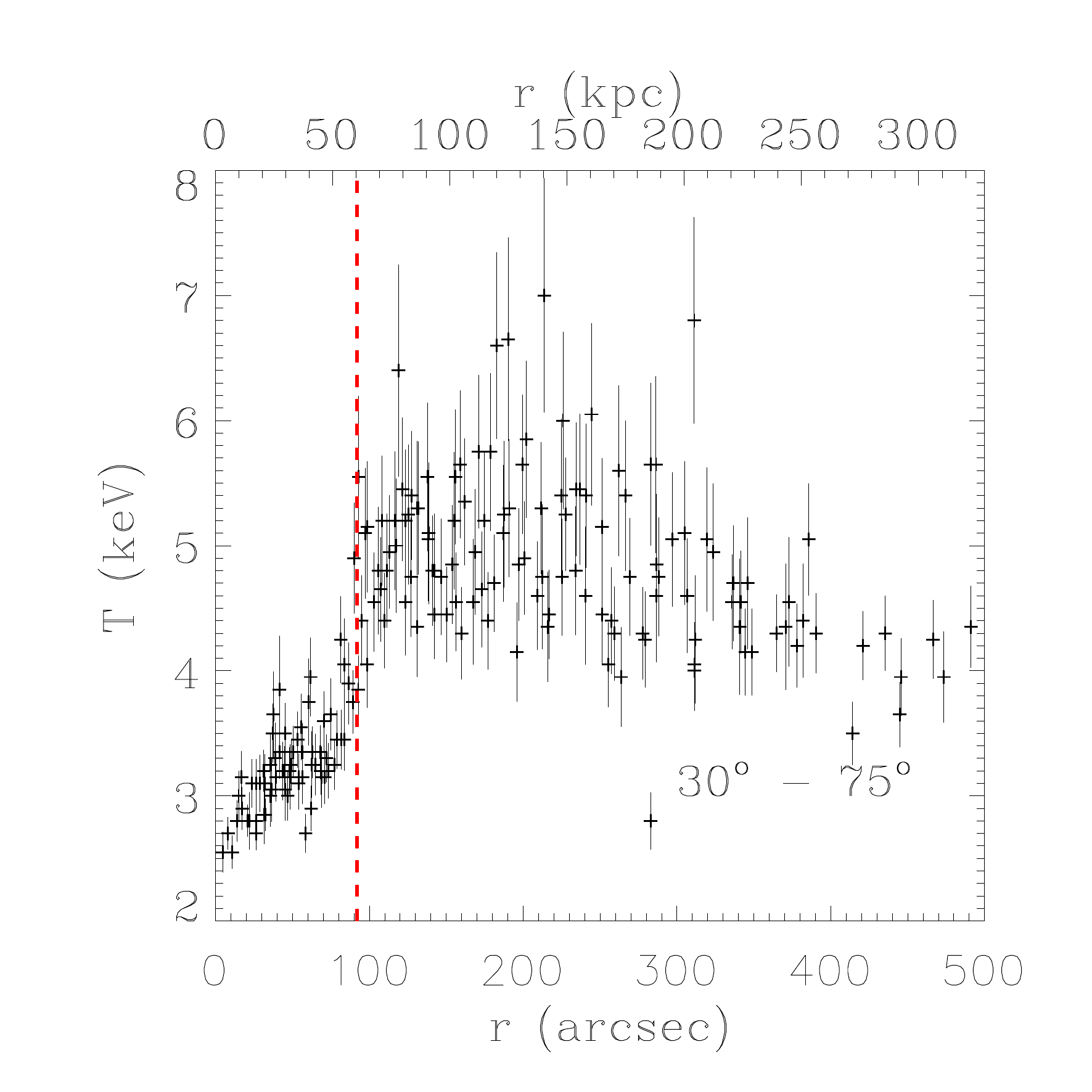}}}
%\hfill
%\centering
%\vspace{5mm}
{{\includegraphics[width=0.25\textwidth]{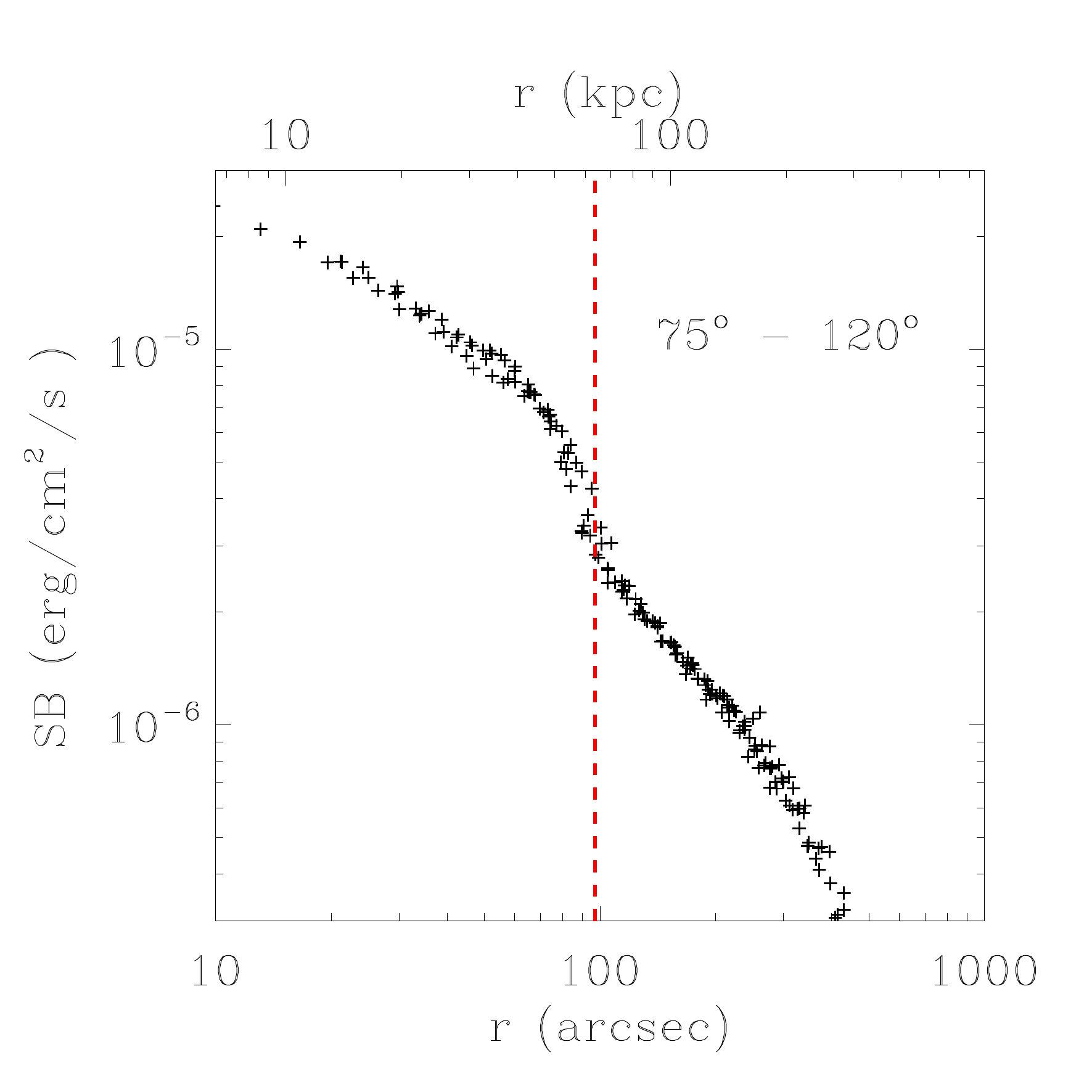}}
%\hspace{5mm}
\hspace{-7truemm}
{\includegraphics[width=0.25\textwidth]{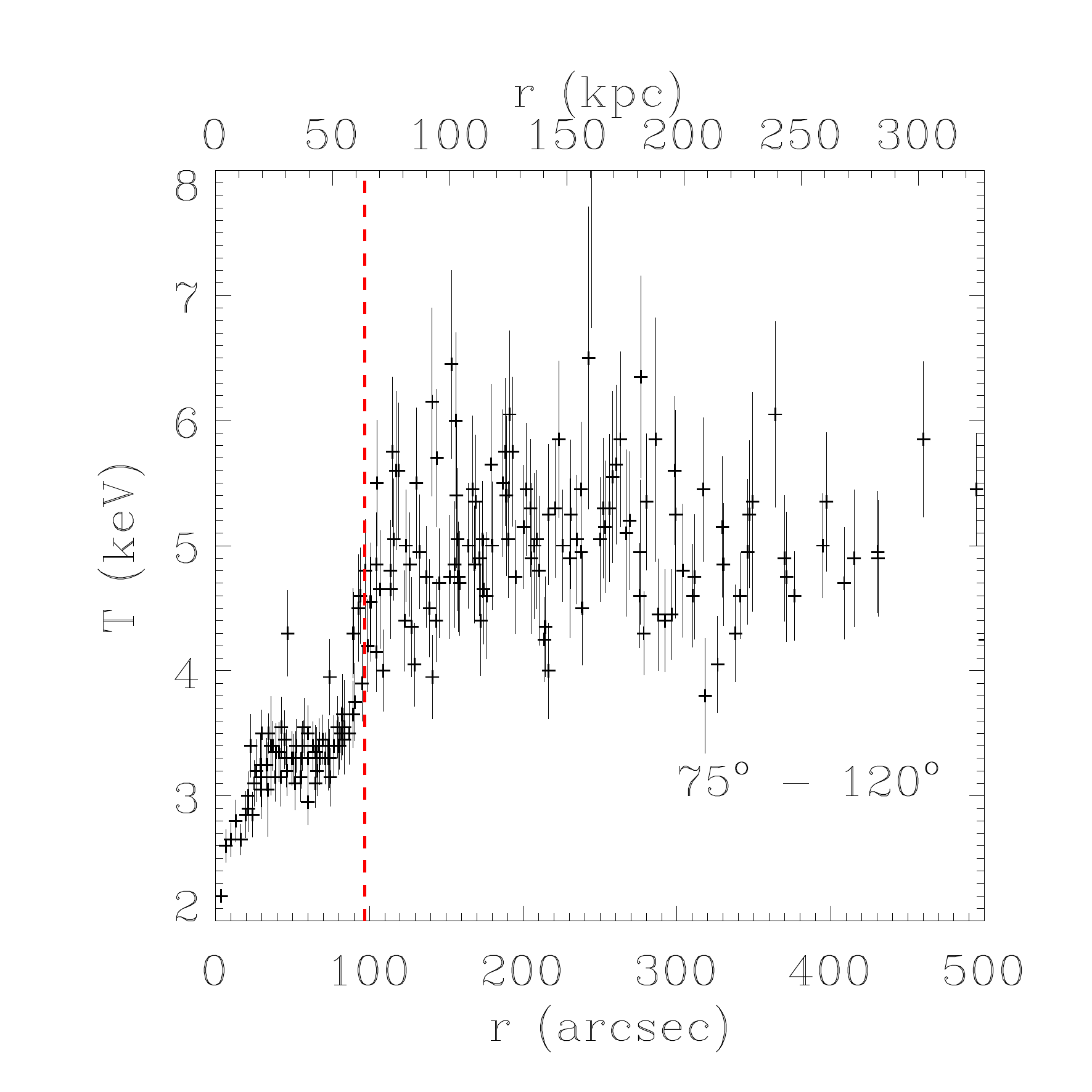}}}
%\hfill
%\centering
%\vspace{5mm}
{{\includegraphics[width=0.25\textwidth]{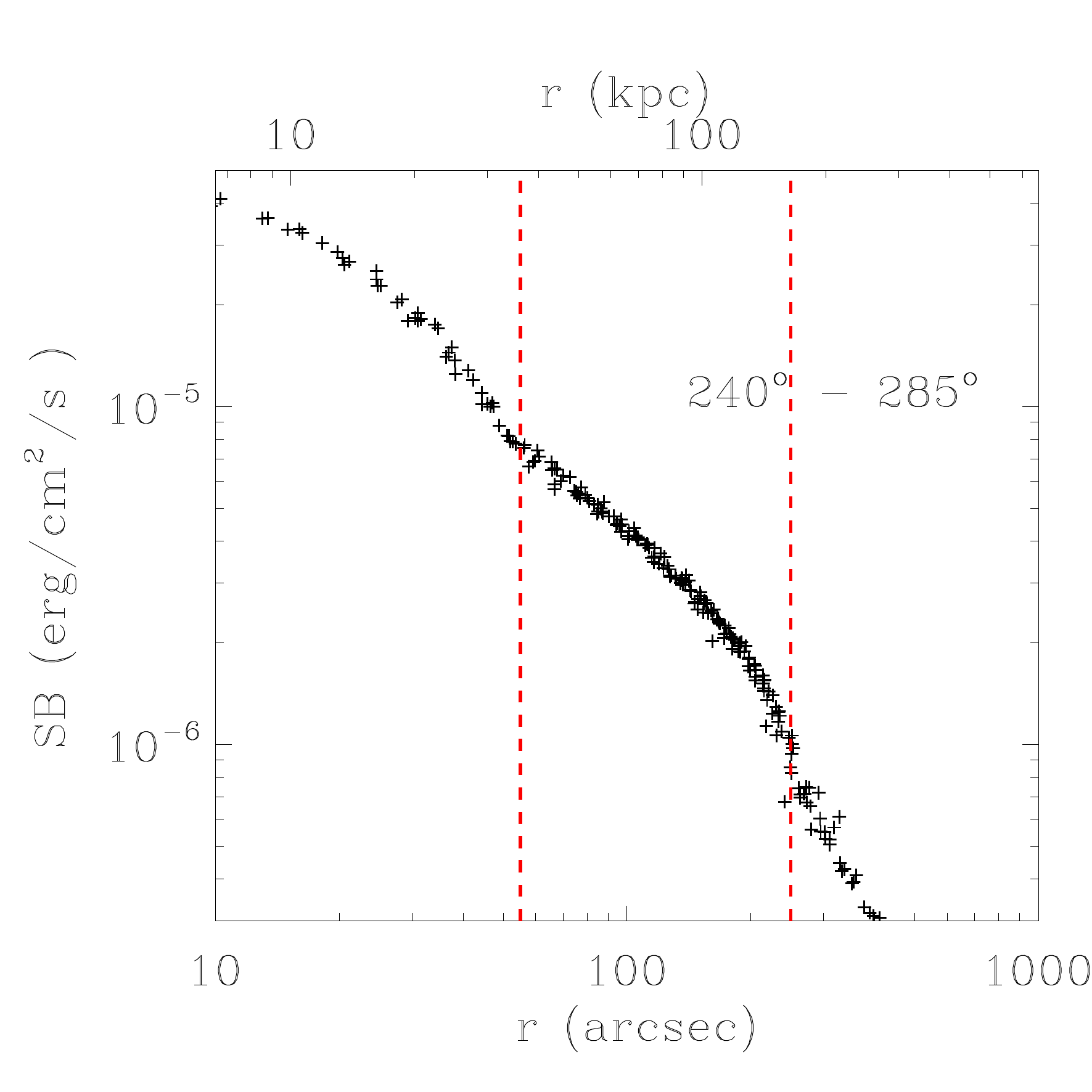}}
%%\hspace{5mm}
\hspace{-4truemm}
{\includegraphics[width=0.25\textwidth]{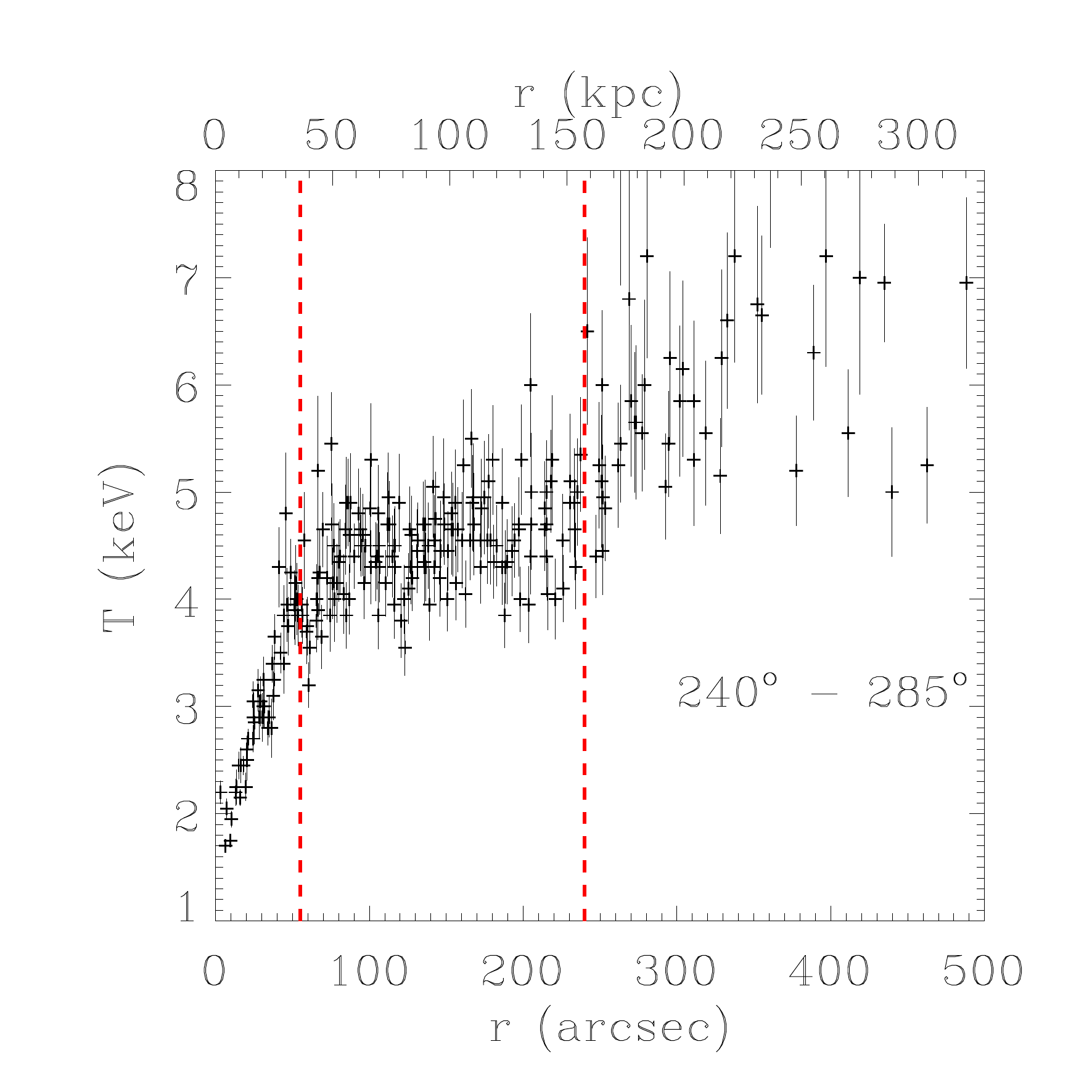}}}
\hfill  
\caption{Surface brightness (left panels) and temperature (right panels) profiles for the cold 
front sectors. 
Profiles are built from the binned maps of Fig. \ref{fig:WVT_mappe_10}. Red dashed lines mark 
the cold fronts positions.}
              
\label{fig:profili_CF}%
    
\end{figure}

The sharpness of the temperature rise across the edges can be better appreciated by looking at the 
temperature profiles of the sectors containing the discontinuities.
We built the temperature profile of each sector of interest by plotting the temperature 
of each map bin belonging to the selected sector versus the
bin distance from the X--ray peak.  
In Fig. \ref{fig:profili_CF} we plot the temperature profiles for the sectors hosting the 
surface brightness discontinuities.
The NNW sector has been split into two subsectors 
[30$^{\circ}$-75$^{\circ}$ and 75$^{\circ}$-120$^{\circ}$] 
because of the boxy appearance of the discontinuity. 
The sector 240$^{\circ}$-285$^{\circ}$ hosts the two southern fronts.
The temperature sharply increases across each edge confirming that all these discontinuities are 
cold fronts. 
Notably, the cluster hosts 4 (when including the E cold front observed by \citealp{Dupke:2003,DW:2007}) 
cold fronts. Notoriously, the development of multiple cold fronts in a relaxed cluster is likely 
to be generated by the sloshing of the central cool gas.

\begin{figure}
  % \centering
\center
   \includegraphics[angle=0,width=9.truecm]{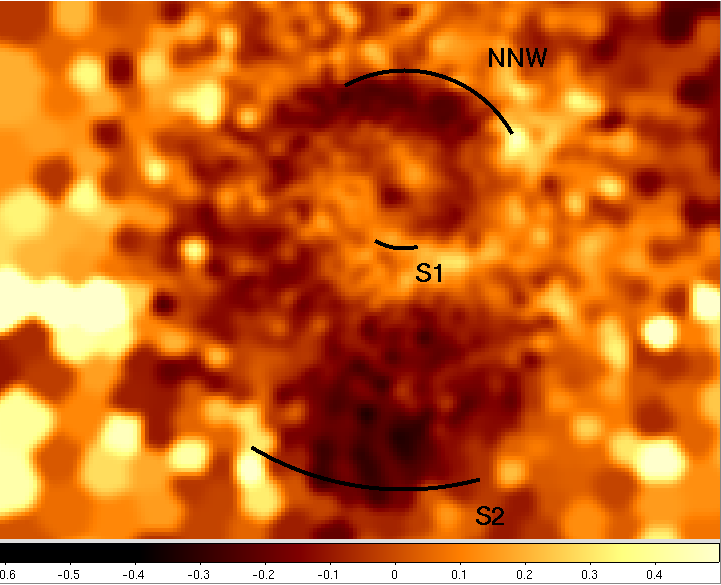} 
   \caption{Entropy residual map. Black arcs mark the detected cold fronts. Residual entropy is zero in
regions whose the entropy equals the averaged entropy; darker regions (negative values) 
have low entropy levels and lighter regions (positive values) have an entropy excess with respect to the average.} 
\label{fig:mappa_residua_K}%
    
\end{figure}

Another marker of the occurrence of the sloshing in A496 is the presence of a spiral 
feature in the temperature and entropy maps. 
Spiral patterns have been observed in the temperature maps of several clusters hosting 
sloshing cold fronts
\citep{Clarke_A2029:2004,Fabian_Perseus:2006,Lagana:2010} 
and are predicted by simulations \citep{AM06,Roediger_ZuHone:2012,R12}.
Numerical simulations by \citet{AM06}
show that the central  cool gas starts sloshing after a minor merger event. 
If the cluster is experiencing an off axis merger, then the gas may acquire an 
angular momentum and the sloshing takes a spiral-like appearance.
Panels (a) and (c) of Fig. \ref{fig:WVT_mappe_10} reveal a hint of such a spiral. 
The spiraling pattern is remarkably sharp in the entropy residual map (Fig. \ref{fig:mappa_residua_K}). 
This map is obtained as $(K -K_{ave})/K_{ave}$ where $K$ is the entropy map and $K_{ave}$ is the
averaged entropy map obtained by averaging entropy in concentric annuli. Hence by definition, regions that in Fig. 
\ref{fig:mappa_residua_K} are zero, correspond to those regions in the cluster whose entropy equals the averaged entropy; 
darker regions (negative values) 
are the cluster regions having low entropy levels and lighter regions (positive values) are the cluster regions having an entropy excess 
with respect to the average.
This method allows to highlight deviations 
from the averaged values so that asymmetries and patterns can be more easily detected.
The entropy residual map shows that 
the central, cool, low entropy gas develops in a spiral-shaped fashion 
extending from the center anti-clockwise towards north where the main NNW cold front is located; 
here the spiral turns eastward and finally expands in the south direction. 
Black arcs in Fig. \ref{fig:mappa_residua_K} mark the cold front positions.
Cold fronts are located along the edge of the spiral and the southern outermost cold front (S2)  
is placed at the tail.

Panel (c) of Fig. \ref{fig:WVT_mappe_10} shows that the gas entropy steeply drops 
in the central regions. This is expected in sloshing clusters. Indeed, simulations by \citet{AM06} 
highlighted that a steep central entropy profile is a necessary condition 
for triggering the sloshing. On the observational side, by analyzing a large sample of clusters 
observed with XMM-Newton, G10 found that clusters that host sloshing cold fronts feature 
entropy drops in their cores, in agreement with theoretical expectations (see however the case of the 
outermost cold front in A2142; \citealp{Rossetti_A2142:2013}).

Unlike the temperature and the entropy maps, the pressure map (panel b in Fig. \ref{fig:WVT_mappe_10}) is regular, 
nearly symmetric, with a mild elongation in the N-NW to S-SE direction. 
Displacements of the pressure from the averaged value are modest, within 10\% in almost all the regions.
This may indicate that the cluster is 
fairly relaxed and that no remarkable perturbations to the underlying gravitational potential
have been induced  by any recent major merger event, as expected in a sloshing scenario,
where the mechanism is thought to be triggered by minor mergers.
The regular configuration of the pressure also confirms that there are no substantial departures 
from hydrostatic equilibrium and that, if some gas is sloshing the motion is subsonic.
This picture is supported by simulations specifically tailored to A496 \citep{R12}. 
The authors reproduce the characteristic spiral pattern and suggest that 
sloshing has been triggered by a minor merger subcluster crossing A496 from the south-west
to the north-north-east, after its passage south-east of the cluster core.

The analysis of the thermodynamical maps highlights that A496 is a sloshing cluster hosting 4 
cold fronts. 
So far, this is one of the few clusters where 4 cold fronts have been detected, together with A2142 
(\citealp{Rossetti_A2142:2013}; Markevitch private communication) 
and Perseus if its western surface brightness excess detected by 
\citet{Simionescu_Perseus:2012} is confirmed to be a cold front. Among this handful of clusters,
A496 is the 
only one hosting 4 cold fronts in a circum-core area ($ r < 250 $ kpc), while A2142 and Perseus 
farthest cold fronts lie at distances $> 0.5$ Mpc.
This characteristic makes A496 a unique cluster to investigate sloshing properties.

\subsection{Characterization of the fronts}
\label{sec:prat}

\begin{figure*}
  % \centering
   \includegraphics[angle=0,width=6 truecm]{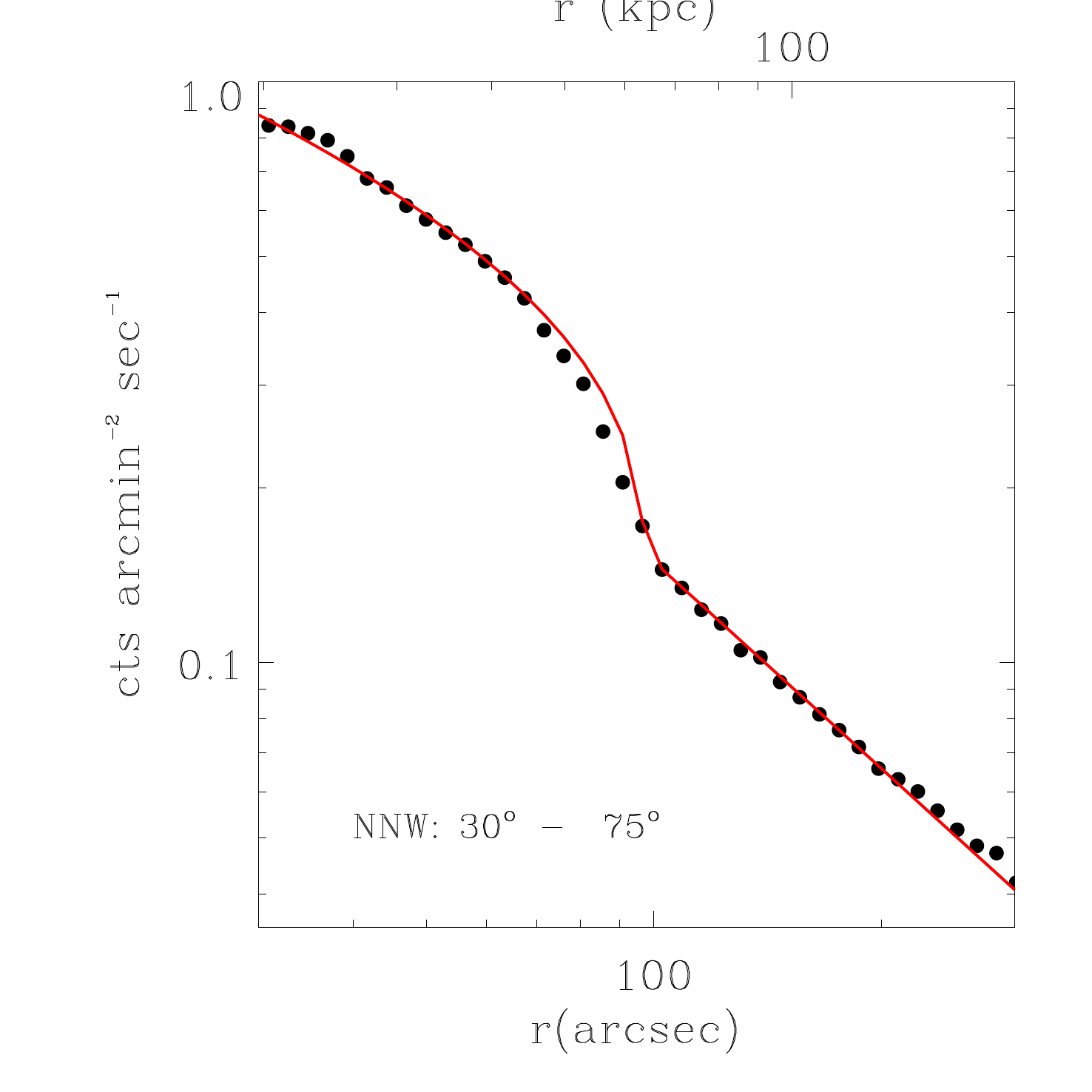} 
   \includegraphics[angle=0,width=6 truecm]{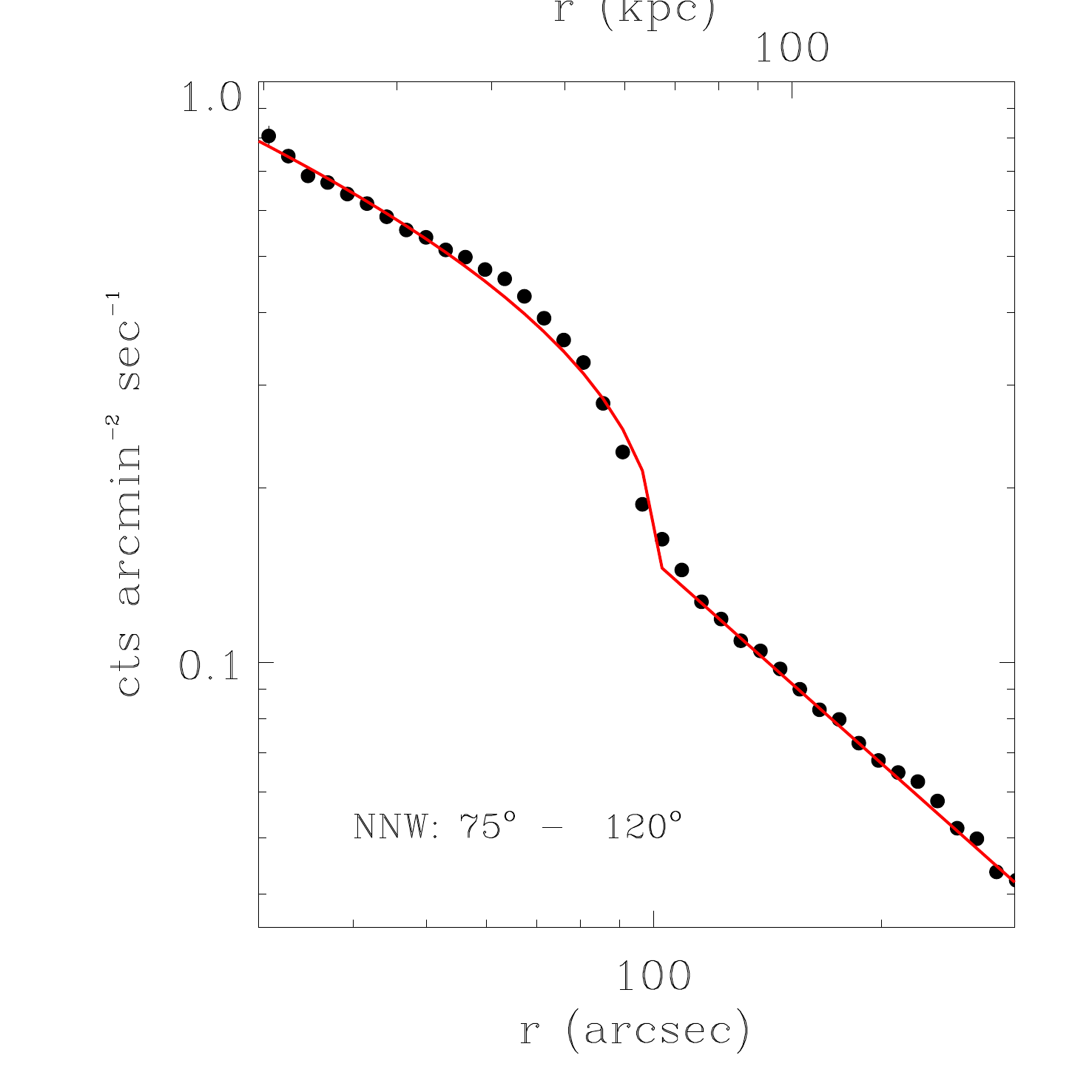}
   \includegraphics[angle=0,width=6 truecm]{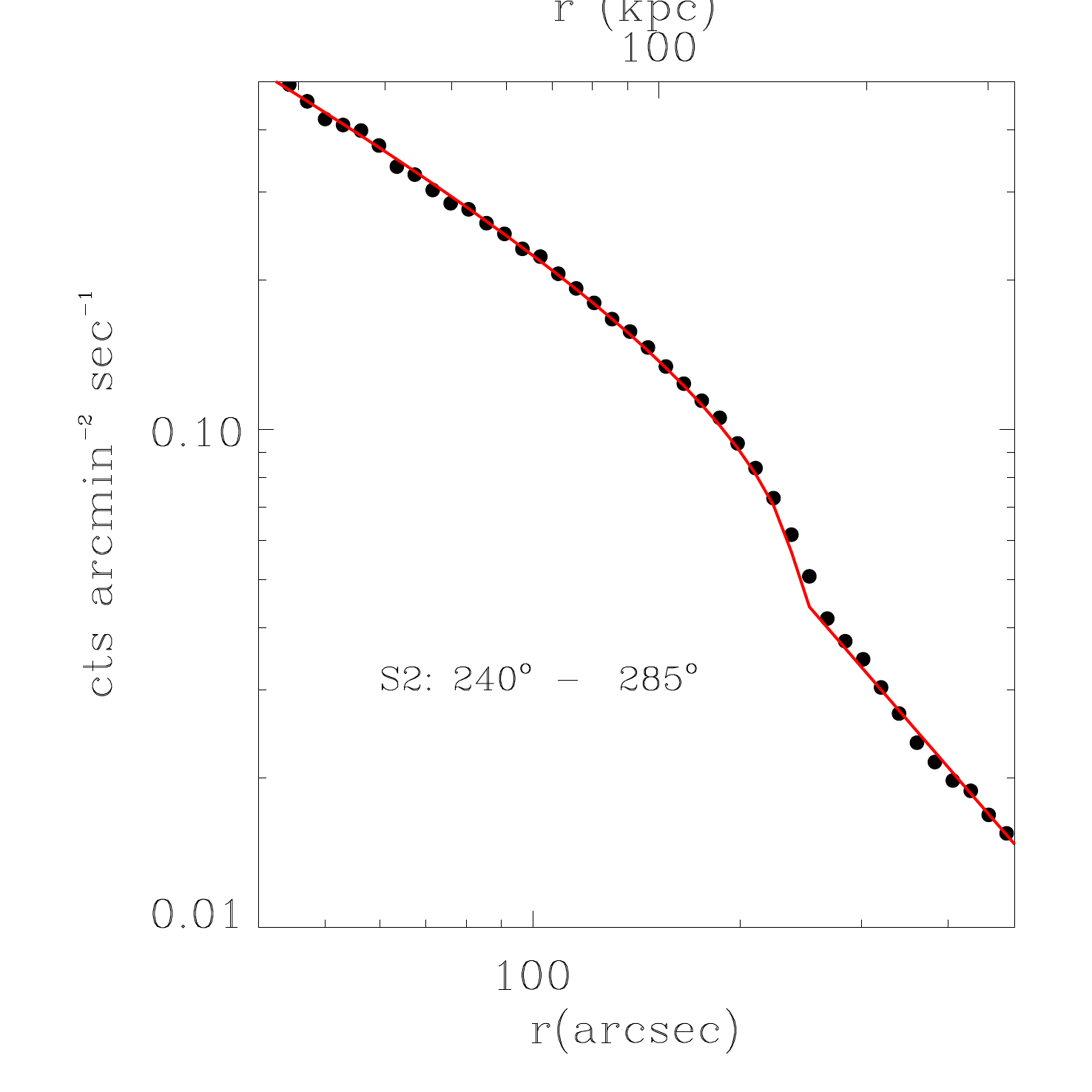}
   \caption{Surface brightness profiles in the three cold fronts sectors $30^{\circ} - 75^{\circ}$ (left panel), 
$75^{\circ} - 120^{\circ}$ (middle panel) and $240^{\circ} - 285^{\circ}$ (right panel). 
The solid red line is the best-fit surface brightness 
obtained assuming that the electron density follows a broken power-law.} 

\label{fig:profili_mach}%
    
\end{figure*}

We characterize our cold fronts by means of a broken power-law (see G10 for details).
The slope and the normalization of each power law are free parameters.
We derive the surface brightness by projecting the emissivity along the line of sight, 
assuming spherical symmetry. We fix the temperatures inside and outside the fronts using 
the spectral analysis results (see Sect. \ref{sec:t_z_prof}).
We fit the observed surface brightness with the projected profiles to derive the power laws 
parameters and find the electron density profile.
By using the density and the temperature, it is then straightforward to derive 
the pressure profile and the pressure jump at the front (see also Appendix \ref{sec:app_press} for 
details).
The innermost eastern cold front E is located at $\sim 16$ kpc ($\sim 24$\as) from the center \citep{DW:2007}, and it is 
too close to the center to be resolved by the XMM-Newton instruments. The
southern S1 cold front lies at $\sim 35$ kpc ($\sim 55$\as) from the center, and though clearly detected by XMM 
a detailed characterization of the profile is difficult.
Thus, we do not analyze these two fronts and focus on the NNW and S2 cold fronts.
We split the NNW cold front into two subsectors because of the boxy morphology of this 
discontinuity.
For the NNW cold front, in the sector 30$^{\circ}$-75$^{\circ}$ we measure a pressure jump 
$p_{in}/p_{out} = 1.40 \pm 0.03$
while in the sector 75$^{\circ}$-120$^{\circ}$ we find a slightly lower pressure jump 
$p_{in}/p_{out} = 1.33 \pm 0.02$.
The front position has been set to 92\as\ and 102\as\ (corresponding to $\sim 60$ and $65$ kpc) 
respectively in the two subsectors.
Finally, for the cold front S2, in the sector 240$^{\circ}$-285$^{\circ}$ we measure a pressure jump 
$p_{in}/p_{out} = 1.28 \pm 0.03$ at the front position 252\as\ (corresponding to $\sim 165$ kpc). %165.37  
In Fig. \ref{fig:profili_mach} we plot the surface brightness profiles of the three sectors 
with their best fits.

According to \citet{Vikhlinin1:2001}, in merging systems, the Mach number and 
the velocity of the cold front can be derived from the pressure jumps.
Pressure ratios of about $1.3 -1.4$, similar to those we have measured, would 
give Mach numbers ${\cal M} \sim 0.6 - 0.7$. Conversely, 
for sloshing cold fronts velocity estimates cannot be derived directly from pressure jumps  
since the cold front moves outwards while the gas is spiraling around the center of the potential well. 
However,
we expect that the pressure ratios provide an order of magnitude estimate of the 
real velocities within a factor of 2. Thus the pressure jumps we detect in A496 imply Mach
numbers in the range 0.3-1.0.

\begin{figure}
\centering
\includegraphics[width=9. truecm]{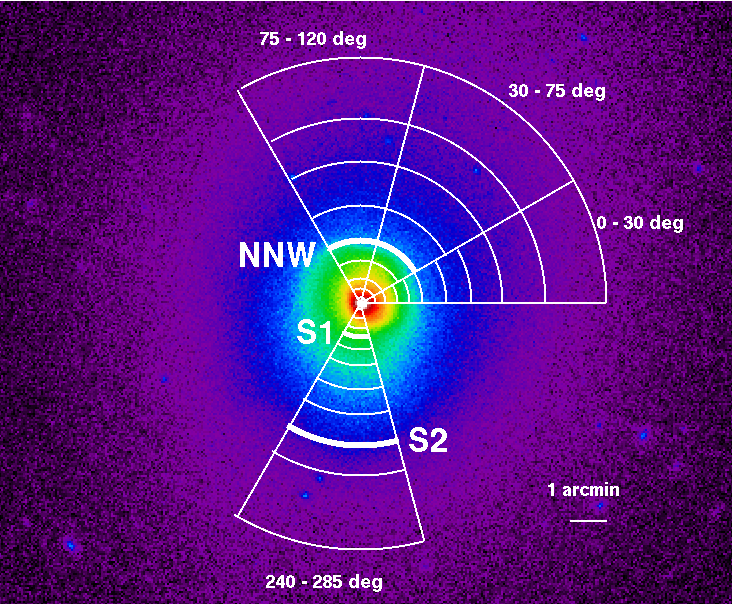}
\caption{X-ray surface brightness image (same as in Fig. \ref{fig:mappa_epic_fx}); white overlay shows regions centered on X-ray peak, used to derive 
profiles presented in Fig. \ref{fig:profili_z}.  NNW, S1 and S2 label the three cold fronts. The fronts positions are marked with a thick arc.}
\label{fig:sect4spec}%
\end{figure}

% SDG NEW FIGURE tempprof dei 4 settori
\begin{figure*}
\centering
{\includegraphics[width=0.48\textwidth]{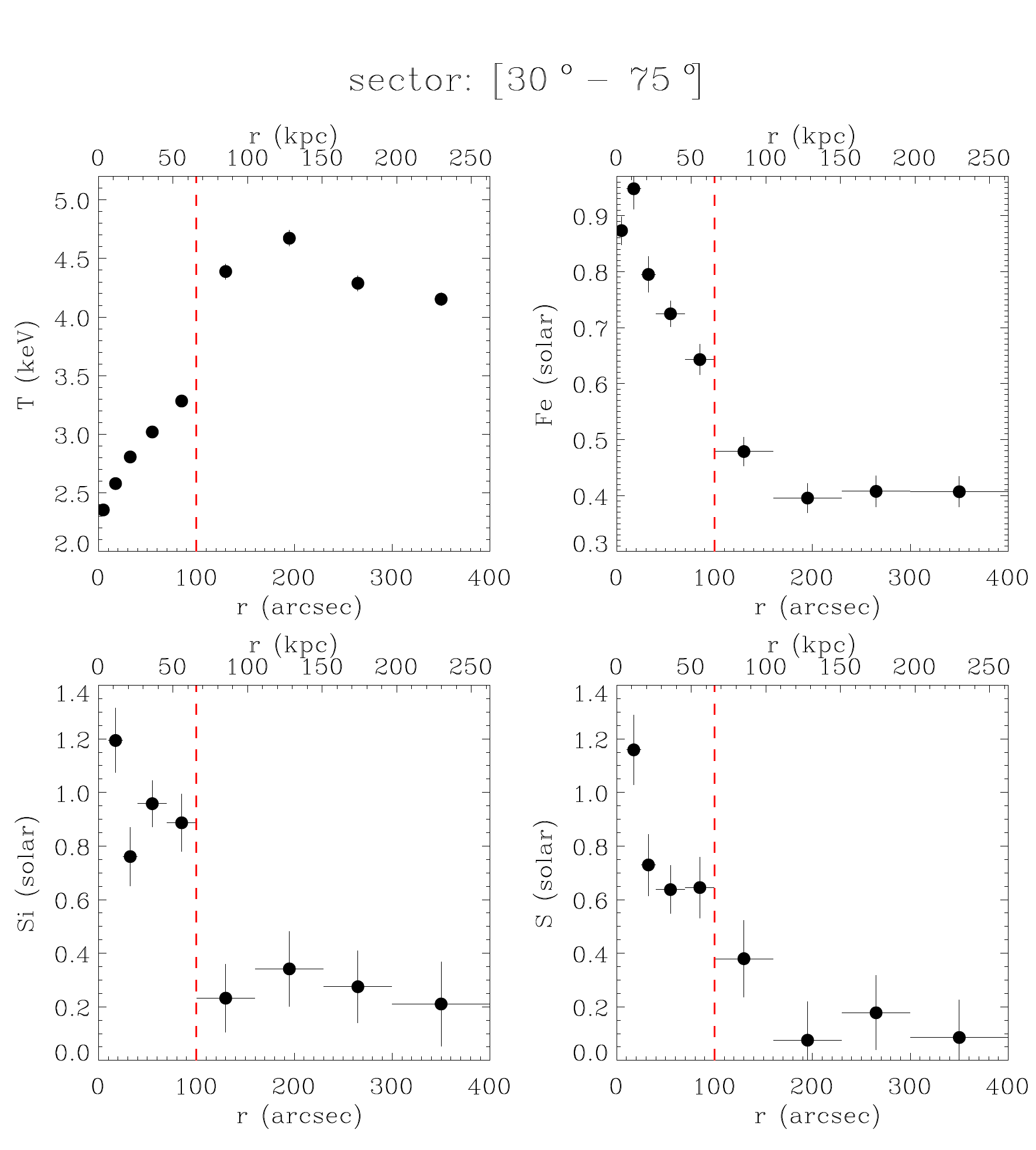}
%\hspace 5mmm
\hspace{5truemm}
{\includegraphics[width=0.48\textwidth]{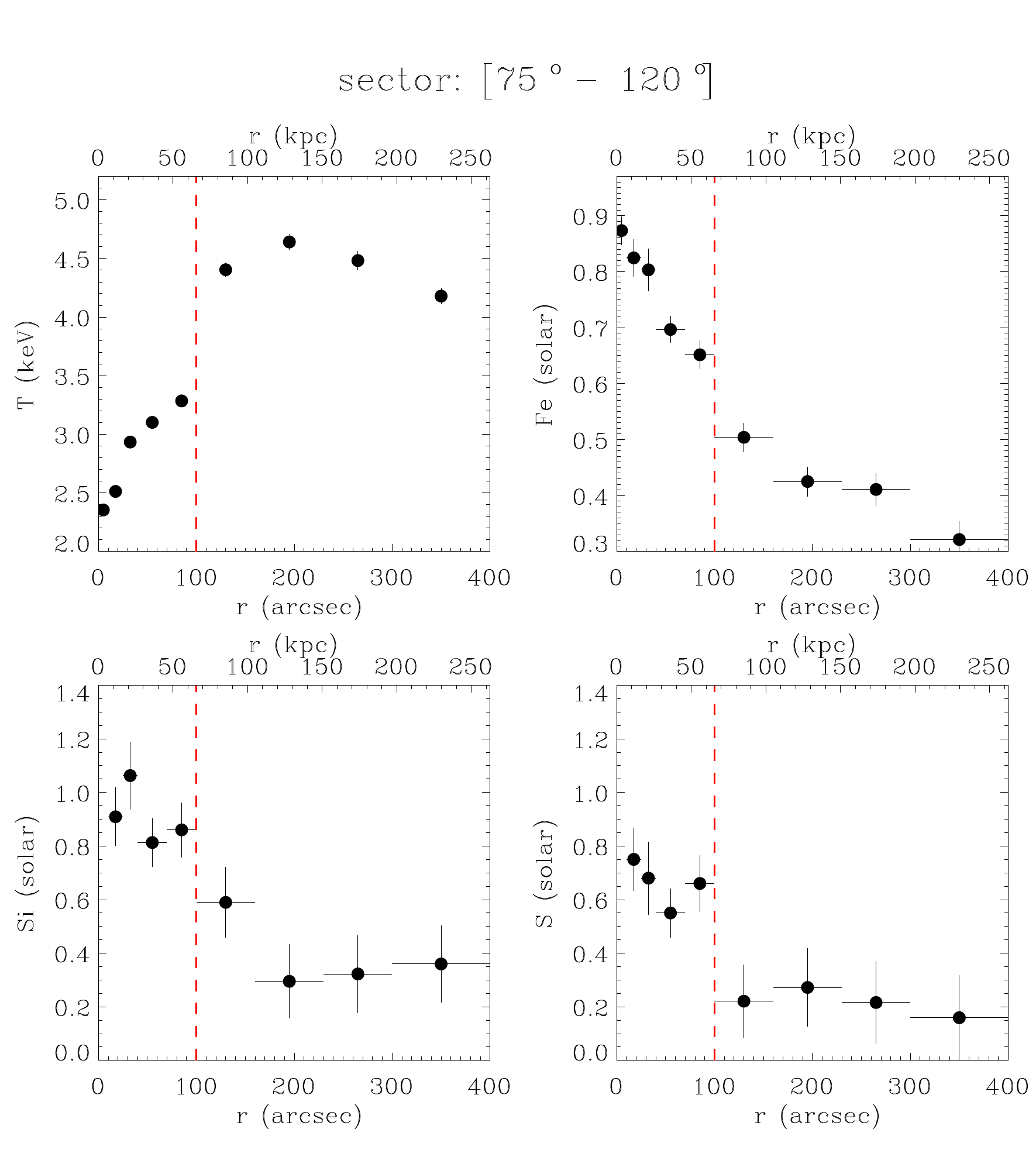}}}
\vspace{1. truecm}
{\includegraphics[width=0.48\textwidth]{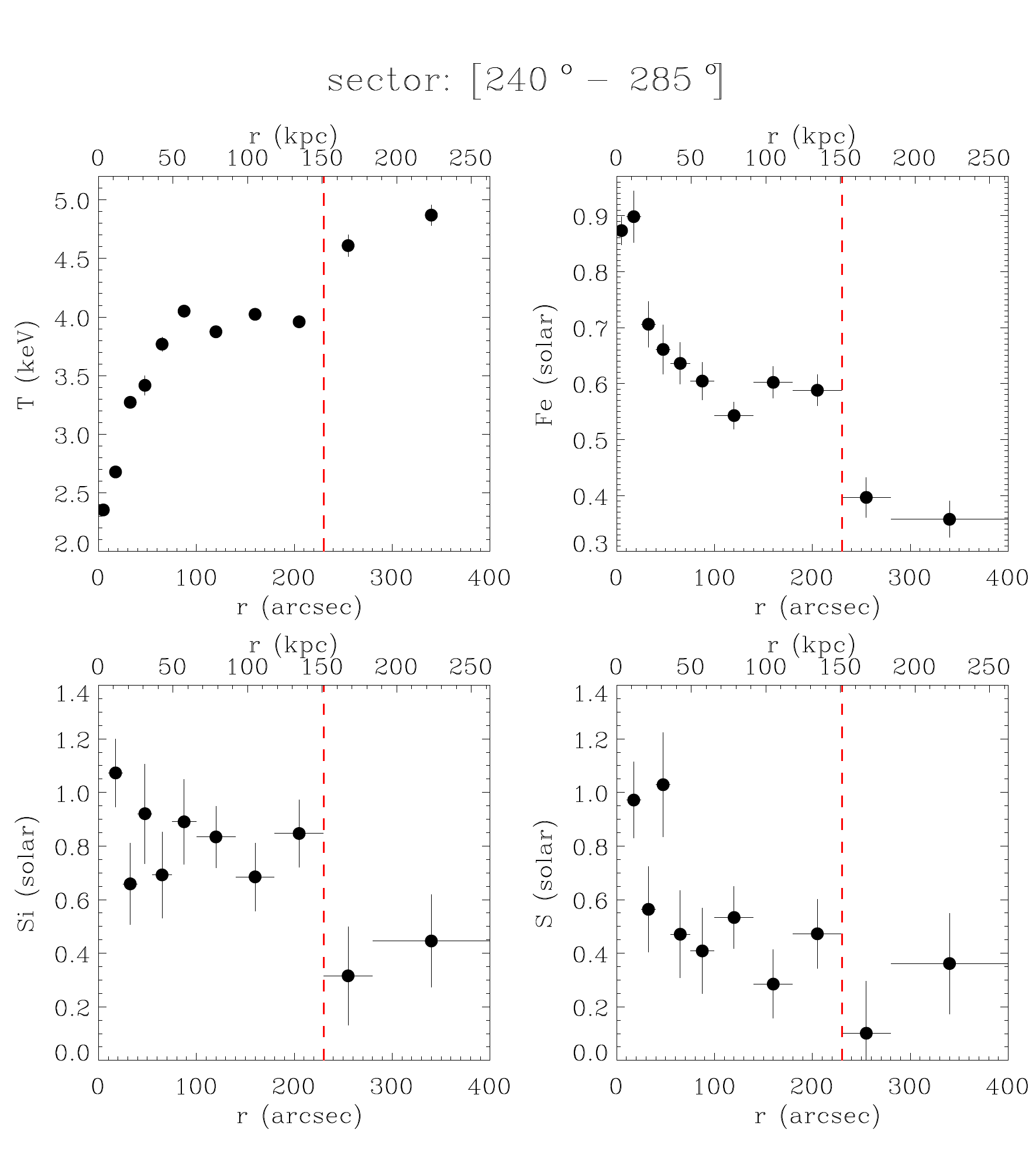} 
%\hspace 5mmm
\hspace{5truemm}
{\includegraphics[width=0.48\textwidth]{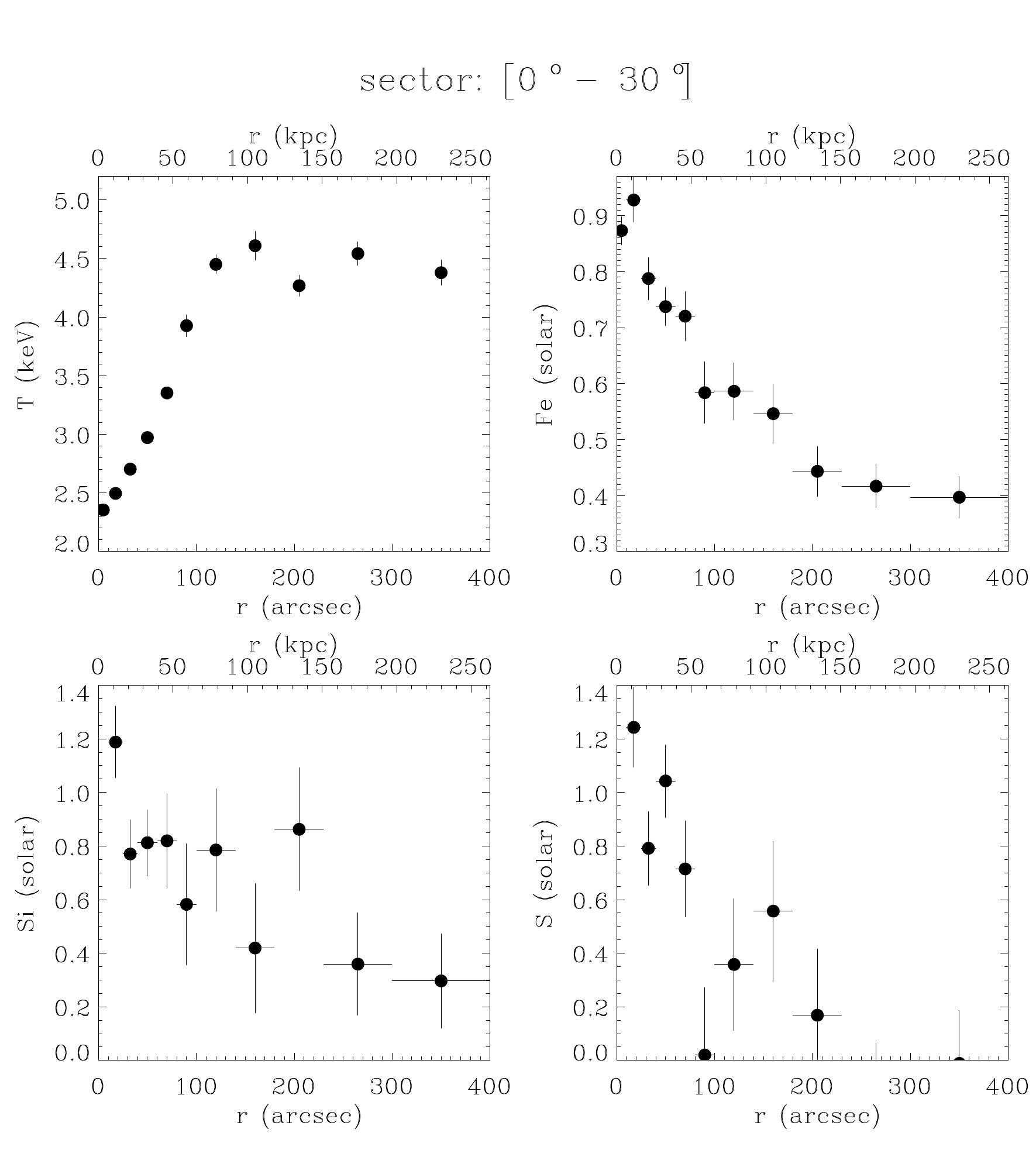}}}
\caption{Temperature, Fe, Si and S abundance profiles for the four sectors $30^{\circ}$--$70^{\circ}$ (top left), 
$70^{\circ}$--$120^{\circ}$ (top right), $240^{\circ}$--$285^{\circ}$ (bottom left) and 0$^{\circ}$--$30^{\circ}$ 
(bottom right). Red dashed lines mark the cold front position. }
\label{fig:profili_z}%
\end{figure*}

\subsection {Radial temperature and metal abundance profiles within sectors}
\label{sec:t_z_prof}
\noindent

Starting from the central peak position (RA 04:33:38; Dec -13:15:41) we divided the cluster, 
in 8 sectors with angles:
0$^{\circ}$, 30$^{\circ}$, 75$^{\circ}$, 120$^{\circ}$, 180$^{\circ}$, 240$^{\circ}$, 
285$^{\circ}$, 330$^{\circ}$ and 360$^{\circ}$ 
(measured from N to E, with 0$^{\circ}$ towards W).
The radial bins in each sector were chosen in order to match the expected cold fronts 
positions. In Fig. \ref{fig:sect4spec} we show, for the sectors of interest, 
the bins used for the spectral analysis.
We extracted spectra for each bin and performed
the spectral analysis with the single temperature (1T) and multi-temperature (GDEM) 
models, as described in Sect. \ref{sec:spec_analysis}.

We first searched for multi-temperature evidences in our spectra.
We found that the only bin where the relative temperature  and Iron abundances of the two models differ, i.e. where
(T$_{\rm 1T}$ -- T$_{\rm GDEM}$)/T$_{\rm GDEM}$ and 
(Z$_{{\rm Fe, 1T}}$ -- Z$_{{\rm Fe, GDEM}}$) /Z$_{{\rm Fe, GDEM}}$ , 
are larger or equal to $\sim$ 10\% and $\sim$ 20\%, respectively, are the ones corresponding to the very 
central circular bin (0--10$^{\prime\prime}$). This is not surprising as in this bin is
included the central AGN.
The other bins 
show relative differences in temperature $\leq 5\%$ and Iron abundances differences are
always consistent within $1\sigma$. 

We concluded that if multi-phase gas is present in these bins, its presence should be 
extremely limited since the mean temperature found by the GDEM model is indistinguishable 
(within the systematics of calibration, spectral analysis or projection errors) from the temperature 
found by the 1T model. 
We refer to the Appendix A for a more extended discussion of the possible presence of 
multi-phase ICM in some regions of A496, in the rest of this Section we will use the results 
of the 1T analysis except for the central bin for which we use temperature and Z$_{\rm Fe}$ 
measured with GDEM.

In Fig. \ref{fig:profili_z}  we plotted the temperature, Iron, Silicon and Sulfur abundance profiles for the three
sectors hosting the NNW (30$^{\circ}$-75$^{\circ}$ and 75$^{\circ}$-120$^{\circ}$) and 
S2 (240$^{\circ}$--285$^{\circ}$) cold 
fronts and, for comparison purposes, a sector (0$^{\circ}$-30$^{\circ}$) where no cold fronts were 
detected. 
Red dashed lines mark the cold front's positions in each panel (whenever cold fronts are present).

All the temperature and Fe abundance profiles shown in Fig. \ref{fig:profili_z} show the typical 
trend of cool core 
clusters, namely a temperature increase, from $\sim 2.5$ keV in the center to $\sim 4.5$
keV in the outermost bins, while the Iron abundance progressively decreases from its central 
peak of $\sim 0.9$ \zsun\ down to an external value of $\sim 0.4$ \zsun.
In Fig. \ref{fig:profili_z} the NNW cold front (split into sectors 30$^{\circ}$--75$^{\circ}$ 
and 75$^{\circ}$--120$^{\circ}$) is clearly marked 
by the temperature jump observed at  100$^{\prime\prime}$ ($\sim$ 65 kpc). 
Interestingly, at the same position the Fe abundance drops abruptly: in both sectors 
at 100\as\ the Fe abundance shows a decrease significant at more than $5\sigma$ 
from the inner part of the front to the outer. 
We note that there is a similar trend, although less significant, also in the Si and S profiles. 
The jumps we measure in Si abundances are significant at least at a $3\sigma$ 
level while jumps in S abundances are significant at a $\sim 2 - 3 \sigma$ level.

A closer look at sector 240$^{\circ}$--285$^{\circ}$ in Fig. \ref{fig:profili_z}, where the S2 cold 
front is located, shows that
the Fe abundance (and possibly also the Si and S abundances) in the region inside the cold front
(with radii in the range between 60--150 kpc) remains roughly constant on a plateau ($\sim 0.6$ \zsun), 
before dropping at the cold front position. 
This area corresponds to the spiral tail (see Sect. \ref{sec:4cf}).

The metal profile in $0^{\circ} - 30 ^{\circ}$ (Fig. \ref{fig:profili_z}) is roughly regular, 
with the metallicity gradually decreasing from the  central peak to the outskirts value. 
The iron abundance profile seems to have some rapid decrease at radius $r \sim 65$ kpc. 
The iron abundance varies from $Z_{Fe}= 0.72 ^{+0.05}_{-0.04}$\zsun\ to $Z_{Fe} = 0.58^{+0.06}_{-0.05}$\zsun. 
This decrease cannot be classified as a discontinuity but we note that its position matches the 
western edge of the spiral 
(see Fig. \ref{fig:mappa_residua_K}). In spite of the presence of the temperature and entropy 
spiral edge, in this sector we do not detect a surface brightness front at the resolution level of XMM-Newton.
It is noteworthy that no surface brightness front 
has been detected in that direction even using Chandra high resolution data \citep{DW:2007}.
A possibility is that the discontinuities are washed out by some Kelvin-Helmoltz (KH) instability 
(see Sect. \ref{sec:KH} and Fig. \ref{fig:sbres}) that smears the front and reduces the jump amplitudes.

\subsection{Metals and spiral pattern}
\label{sec:spiral}

Both the metal discontinuity across the fronts (which match the spiral edges) and the metal 
excess in the wide region 
of the spiral tail point to a connection between the spiral and the areas with high metallicity.
We have performed a spectral analysis to investigate this possible correlation. 

\begin{figure}
   \centering
   \includegraphics[width=9 truecm]{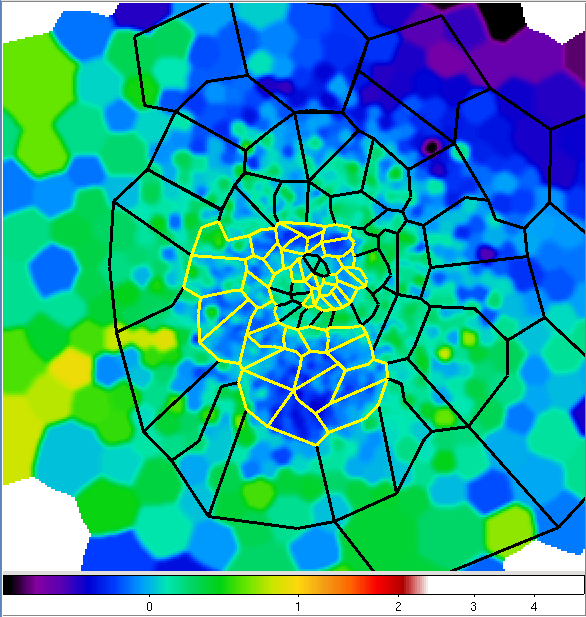}  
 \caption{Entropy residual map with polygons used for the spectral analysis. 
Yellow labels the IN polygons and black labels the OUT polygons. The entropy map is the same of 
Fig. \ref{fig:mappa_residua_K} with the color scale adjusted 
so that blue regions have negative residual entropy values and green regions have positive residual
entropy values.}
              \label{fig:spiral_in_out}%
\end{figure}

We use the entropy residual map (Fig. \ref{fig:mappa_residua_K}) to distinguish between polygonal
regions lying on the spiral pattern and polygonal regions outside
the spiral.
We built polygons for spectral analysis by grouping the polygons obtained from the WVT binning. 
The grouping follows the spiral 
pattern and maintains the S/N roughly constant. The extracted spectra typically 
have 6000-10000 counts in each MOS and
12000-20000 counts in pn.
Each spectral polygon is classified as a polygon lying inside the spiral (IN) or outside the spiral
(OUT).  
Spectral polygons are shown in Fig. \ref{fig:spiral_in_out}: yellow marks the IN polygons and 
black marks the OUT polygons.

\begin{figure}
   \centering
\includegraphics[width=9.truecm]{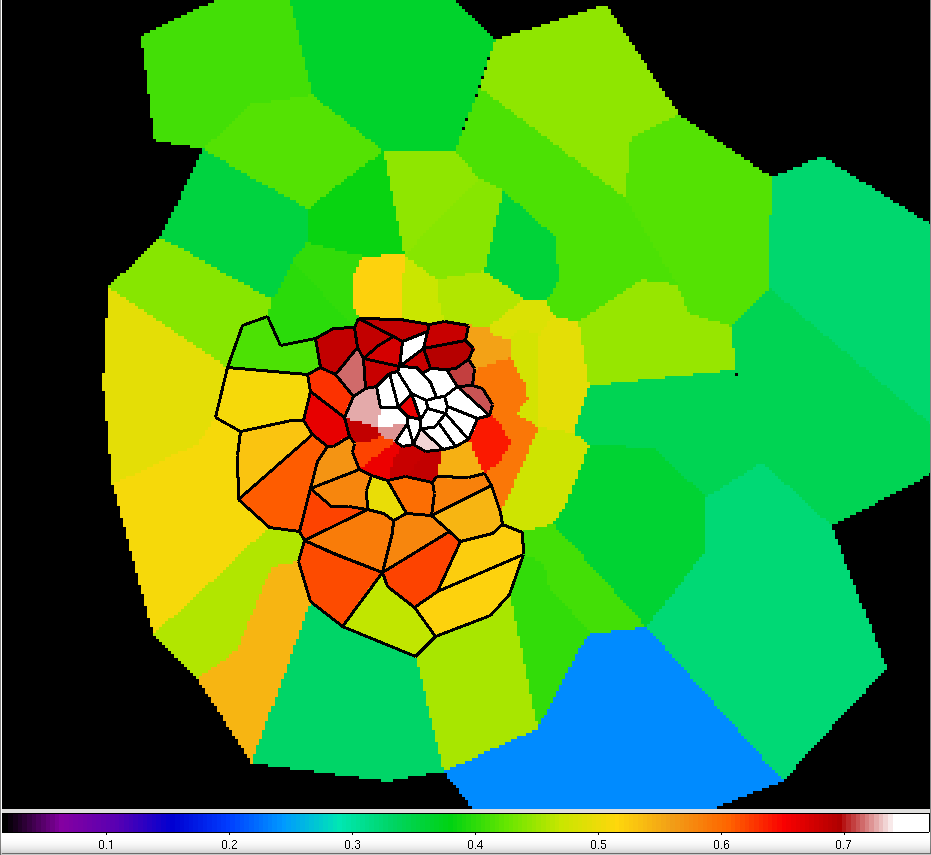} 
 \caption{Iron abundance map obtained from spectral analysis. The color bar indicates the
 metal abundance in solar units. IN polygons are overlaid.}
              \label{fig:Z_Zres_midi}%
    
\end{figure}

As outlined in the previous section, the ICM spectra are well described by a 1T model except for the central bin 
(see also  Appendix \ref{sec:spec_models}
for the details). Hence, spectra extracted from the polygons are fit with a 
single temperature {\em vapec} model. 
Comparison of spectrally measured temperatures  with 
those obtained using the WVT+BBT algorithm shows good agreement.
Fig. \ref{fig:Z_Zres_midi} shows the 
metal abundance map. The spiral pattern is visible. High metallicity polygons (white and red in the 
map color scale) 
follow the low entropy spiral configuration showing that the metal rich gas does lie on the spiral.

The correlation between the low entropy and the metal excess, may be understood under the assumption 
that prior to the onset of the sloshing the metal rich low entropy 
gas was located at the bottom of the potential well, i.e. at the center of the system. 
After the gas is set into motion, presumably by a perturbing subsystem, the low entropy 
metal rich gas moves outwards generating cold fronts and eventually forming the spiral structure.
The difference in entropy and metallicity between 
the gas in and outside the spiral indicates that processes mixing the 
lower entropy higher metallicity gas with the ambient medium are relatively ineffective.

The metallicity map (Fig. \ref{fig:Z_Zres_midi}) also highlights the presence of a high metallicity 
region in the SE part 
of the cluster (orange tones in the map color scale). This point will be discussed later
(see Sect. \ref{sec:SE_excess}).

\begin{figure}
   \centering
  \hspace{-1.1 truecm}
   \includegraphics[angle=0,width=9.5 truecm]{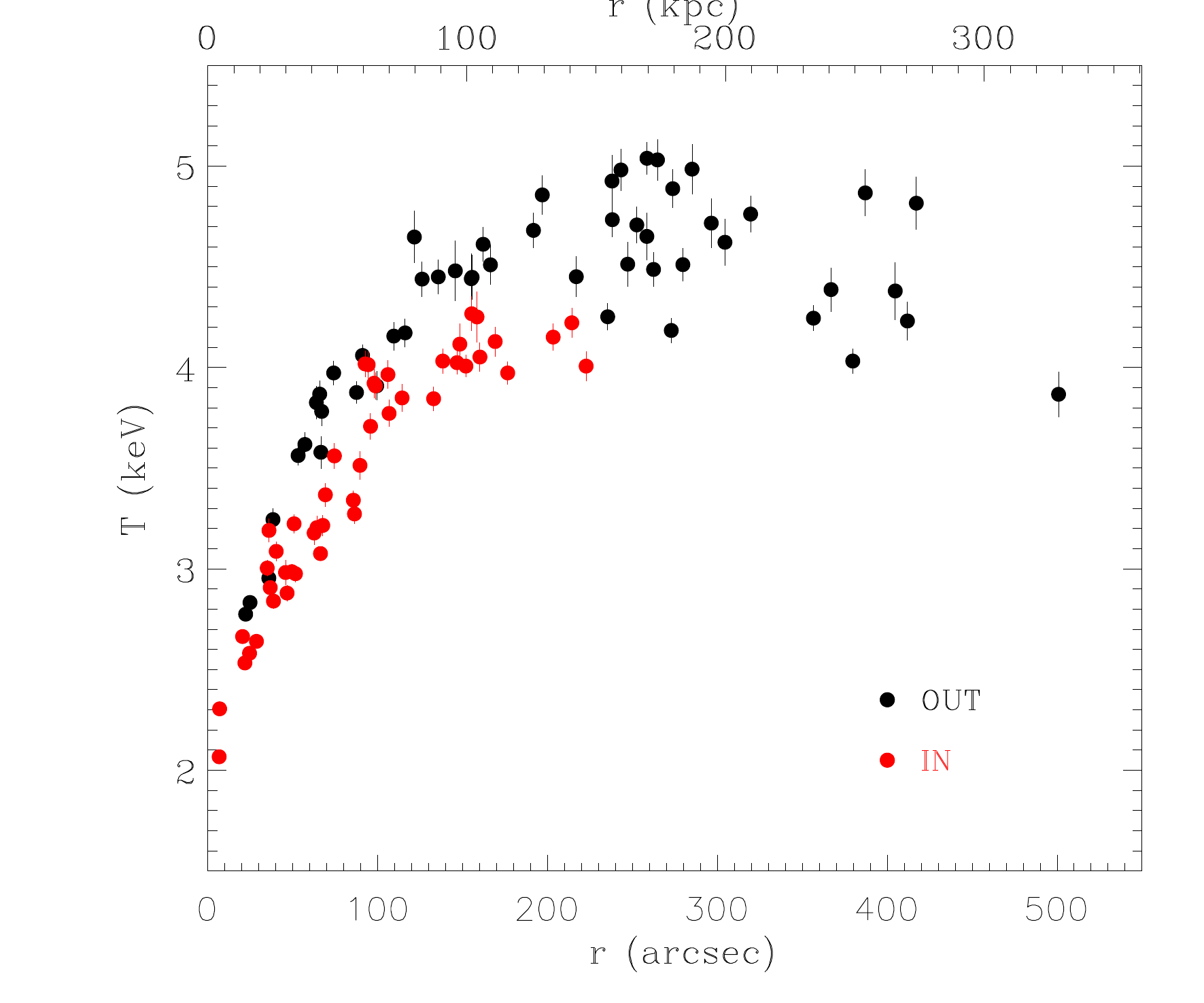}  
 \caption{Temperature profile derived from spectral analysis in the polygonal regions. 
 Red marks the IN polygons, 
black marks the OUT polygons.}
              
              \label{fig:Tprof_midi}%
    
\end{figure}

\begin{figure}
  \centering
  \hspace{-1.1 truecm}
   \includegraphics[angle=0,width=9.5 truecm]{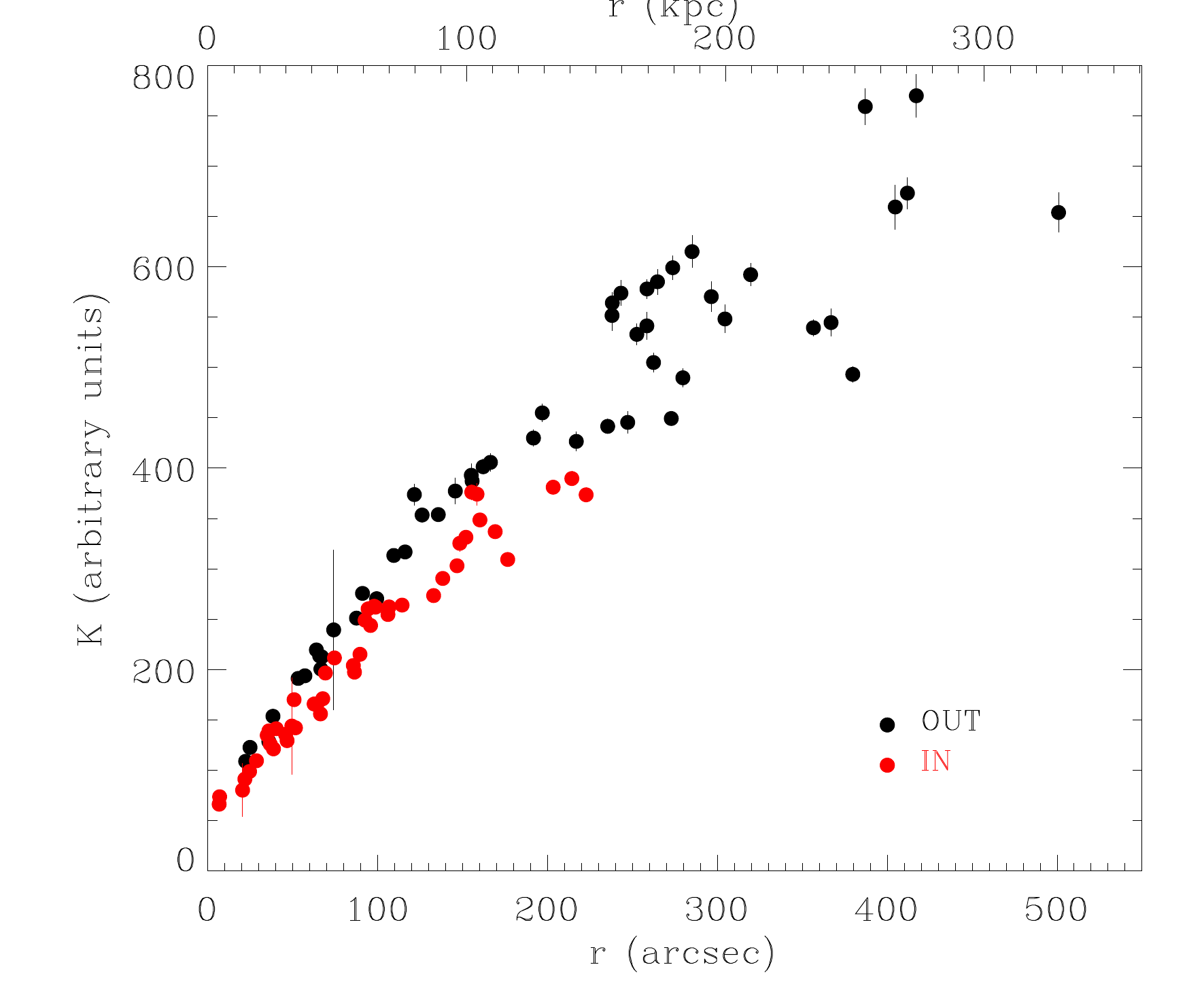}  
 \caption{Entropy profile derived from spectral analysis in the polygonal regions. 
 Red marks the IN polygons, 
black marks the OUT polygons.}            
  \label{fig:entprof_midi}%
    
\end{figure}

\begin{figure}
   \centering
   \hspace{-1.1 truecm}
   \includegraphics[angle=0,width=9.5 truecm]{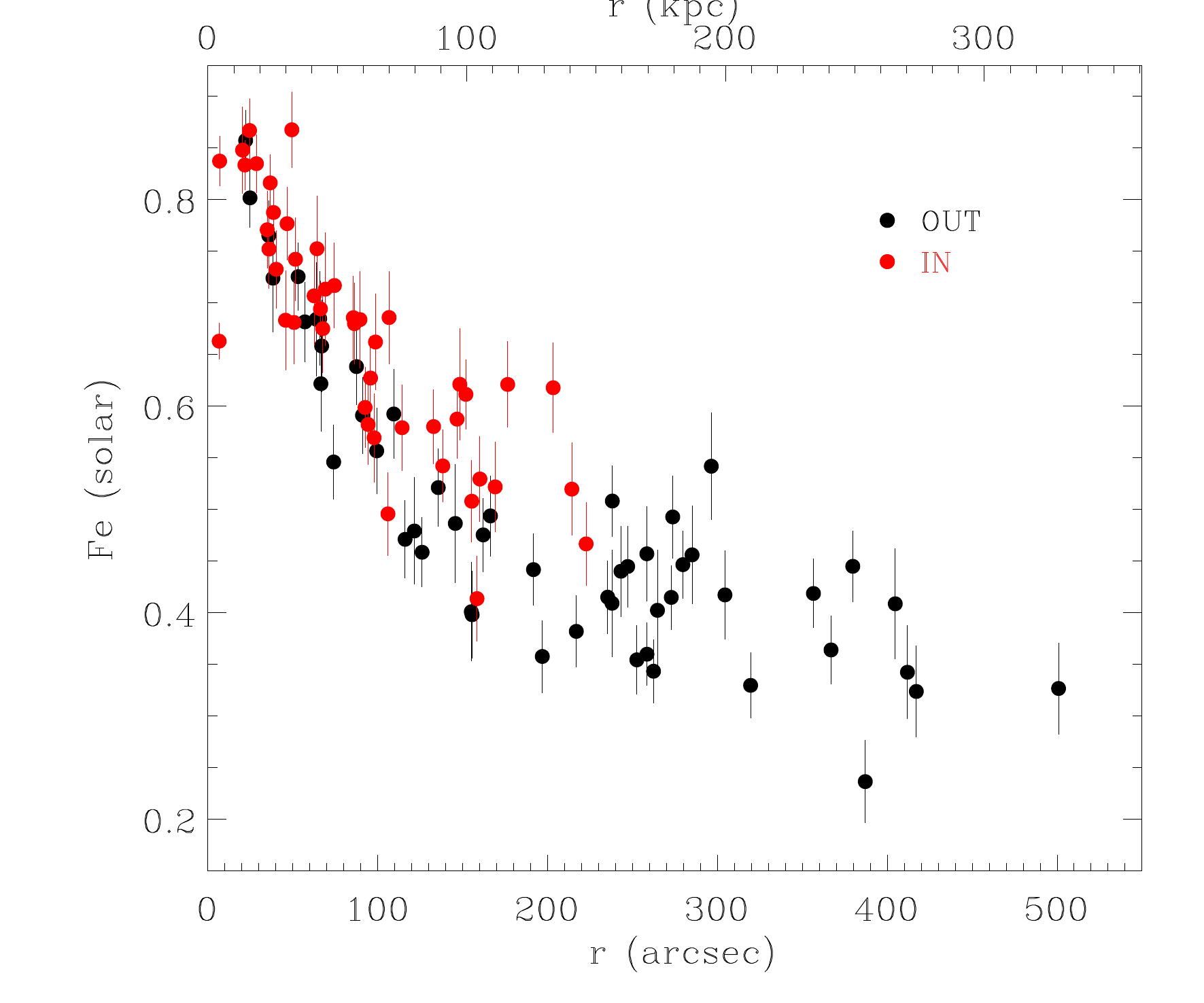}  
 \caption{Fe abundance profile derived from spectral analysis in the polygonal regions.
 Red marks the IN polygons, 
black marks the OUT polygons.}
              
   \label{fig:feprof_midi}%
    
\end{figure}

For a better visualization of the differences between the properties of the gas on the spiral and 
out of the spiral, 
we plot the thermodynamical quantities of each polygon as functions of its distance from the X-ray peak. 
Temperature and entropy profiles are reported respectively in Fig. \ref{fig:Tprof_midi} and 
Fig. \ref{fig:entprof_midi}, red points mark 
the IN polygons and black points mark the OUT polygons. 
We recall that we use projected quantities as explained in detail in Sect. \ref{sec:xmaps}.
For a given radius, the entropy and the temperature are generally lower in the IN regions than in 
the OUT regions. 
Fig. \ref{fig:feprof_midi} shows the iron abundance profiles. 
The metal abundance of OUT polygons follows the typical behavior of a cool core cluster
with the metallicity peak 
in the center and the decreasing trend towards the outer regions.
The metallicity of the IN polygons has a similar trend but offset high with respect to the 
one measured for OUT polygons.
The gas lying on the spiral tail (red points in the radial range $120$\as\ - $240$\as\ 
corresponding to $\sim 80 -160$ kpc) 
has a metal abundance $Z_{Fe} \sim 0.6$ \zsun, which is lower than the peak value $Z_{Fe} \sim 0.8-0.9$ \zsun\ 
and similar to the one observed some  $\sim 80$\as\ - $100$\as\  ($\sim 50-65$ kpc) from the 
center in other azimuthal directions. 
Under the assumption that the metal mixing is inefficient this would imply that the gas on the 
spiral tail 
does not come from the center of the cluster but from an intermediate position, $\sim 50-65$ kpc
from the center.


\section{Discussion}
\label{sec:discussion}

\subsection{Iron abundance and sloshing}
\label{sec:fe-slosh}
The presence of multiple cold fronts arranged in a spiral pattern is a clear 
signature of the onset of the sloshing mechanism in the central regions of the clusters.
Our results show that the Iron abundance sharply drops across the detected cold fronts
(see Fig. \ref{fig:profili_z}).
A less significant drop is also seen in Silicon and Sulfur. 
This metals discontinuity across the front can be promptly explained within the sloshing scenario: 
when the minor merger triggers the sloshing mechanism, the low 
entropy central gas is displaced from the center and pushed out generating the contact discontinuity, 
i.e. the front.
This dense, and cool gas comes into contact with the less dense, hotter ambient gas and
this further enhances the contrast in density (and surface brightness) and 
temperature. Assuming that prior to the onset of the sloshing process the core of A496
was characterized by a metal abundance excess, a metal discontinuity will be created along with the surface 
brightness and the temperature jumps.

The metallicity drop across the discontinuity is observed in several sloshing clusters
(e.g. Perseus: \citealp{Fabian_Perseus:2011}; Centaurus: \citealp{Sanders_Centaurus:2006}; 
A2204: \citealp{Sanders_A2204:2005, Sanders_A2204:2009}).
In addition to the metal abundance discontinuity across the front, A496 has a notable 
characteristic concerning the metal distribution:
high metallicity regions are arranged following the same spiral feature 
traced by the low entropy gas, as apparent in Fig. \ref{fig:Z_Zres_midi}. This is highlighted in 
Fig. \ref{fig:feprof_midi}, where regions located on the spiral (red points) have a higher metal 
abundance than the ambient gas (black points) lying at the same distance from the center.
The most straightforward interpretation is
that the sloshing process allows the low entropy
metal rich gas located at the center to be moved
outwards. The spiral shape is likely due to the
presence of angular momentum in the ICM,
transferred from the subcluster to the core gas during the off-axis merger \citep{AM06,ZuHone:2011}.

Remarkably, the discrepancy between the metallicity of the gas lying on the spiral 
and of the environment gas lying at the same radius increases with distance: 
the gas lying on the tail of the spiral,  $\sim 100-150$ kpc from the center has a 
metallicity $\sim 0.6$ \zsun, significantly higher than the averaged metallicity of the 
hotter surrounding medium ($ \sim 0.4$ \zsun).
This means that mixing processes are not very efficient.
Indeed if mixing were efficient, the cool gas would gradually reduce its metal abundance while sloshing and will progressively 
tend to homogenize with the surrounds. In this scenario, the gas 
currently residing in the spiral tail would be initially located in the very central regions (say $\simlt 50$ kpc) where 
environment metallicity is $ \simgt 0.6$ \zsun; the higher the mixing efficiency the higher the starting metal content.
Though the mechanism would involve all the gas of the spiral, the gas of the tail would cover the largest distance 
and would undergo the mixing process for a longer time. Hence in presence of efficient mixing processes,
the gas of the tail is expected to experience a higher level of mixing and reach a higher degree of homogenization with the ambient gas.
This is in conflict with results shown 
in Fig. \ref{fig:feprof_midi} where the largest discrepancy between the IN and OUT gas is observed in the tail.

In the hypothesis of the absence of mixing, we can therefore infer that 
the gas located on the spiral tail did not originate from the 
the center but probably comes from an intermediate region, $\sim 50-60$ kpc. This is in agreement 
with simulations 
by \citet{AM06} (see Fig. 8 in their paper), who show that the relative radial displacement $r/r0$, 
(i.e. the 
ratio between the final distance and starting distance from the center) of the sloshing gas,
is about 1.5-2.

\subsection{Cold fronts velocity and age}
\label{sec:age}

In sloshing systems, the cool gas moves subsonically in the ambient ICM, with relatively low Mach number values.
Although the velocity of the front cannot be directly obtained from pressure ratios in these systems 
(see Sect. \ref{sec:prat} for details),
the pressure jump can provide a rough estimate (within a factor of 2, \citealp{Vikhlinin:2002})  of the real velocity and, 
consequently, of the 
cold front age.
Assuming a Mach value  $ \simeq 0.5$, which corresponds to a 
cold front velocity of $\sim 500 $ km/s, in the hypothesis of an averaged ICM temperature $\sim 4$ keV,
the front would cover 
the distance of  $\sim 150$ kpc from the center in about $\sim 0.3$ Gyr.

Simulations by \citet[see their Figs. 3 and 7]{AM06} show that the time necessary for the central gas to turn back 
and wrap around to form 
the central ring of the spiral and for the outermost gas to form the spiral tail is typically $0.6-0.7$ 
Gyr after the subcluster pericentric passage.

Consistently, \citet{R12} developed high resolution hydrodynamical simulations tailored
to reproduce A496 history to study 
the evolution of the cold fronts in this cluster and 
they find that the sloshing for A496 is likely to be induced by a perturbing subcluster which 
crossed from the SW to the NNE,
$0.6-0.8$ Gyr ago.

\subsection{Kelvin-Helmoltz instabilities}
\label{sec:KH}

\begin{figure}
\centering
{\includegraphics[width=0.5\textwidth]{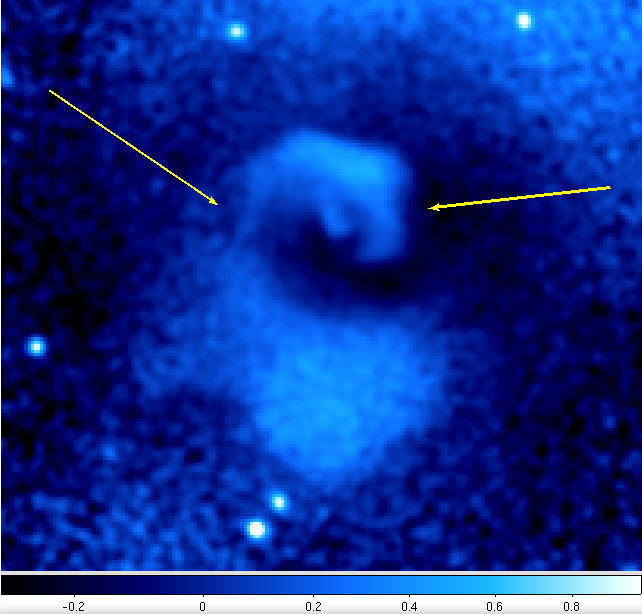}} % 0.45 in non-referee format
\caption{Surface brightness residual map. Arrows mark the regions where the spiral edges are 
smeared out, possibly by Kelvin-Helmholtz instabilities.}
\label{fig:sbres}
\end{figure}

The spiral pattern drawn by the cool, low-entropy gas is also clearly visible in the residual surface 
brightness map (see Fig. \ref{fig:sbres}). The figure reveals a noteworthy feature: 
the spiral portion east of the 
center connecting the main NNW cold front and the tail appears to be significantly thinner than 
elsewhere.
The same occurs on the spiral section west of the center. 
These regions are marked by yellow arrows in the figure. The edges of the spiral there, 
are smeared out so that the 
spiraling feature is less sharp.
A similar configuration can be seen in \citet{AM06} maps (see their Fig. 7, left bottom panel).
This smearing of the front could be explained by the presence of Kelvin-Helmoltz (KH) instabilities.
Although cold fronts are stable phenomena, shear flows can be strong enough to trigger KH instabilities.
These instabilities tend to create kinks and distortions in the smooth arc-like profile of the 
fronts and of the spiral.
The A496 spiral is characterized by features that are similar to those produced by KH instabilities in simulations 
(see \citealp{R13,ZuHone:2013}). 
The elongated shape of A496 probably enhances these features as the elongation 
stretches the spiral.

\subsection{Entropy and metal abundance correlation}

\begin{figure}
 
\centering
\hspace{-1.1 truecm}
\includegraphics[angle=0,width=9.5 truecm]{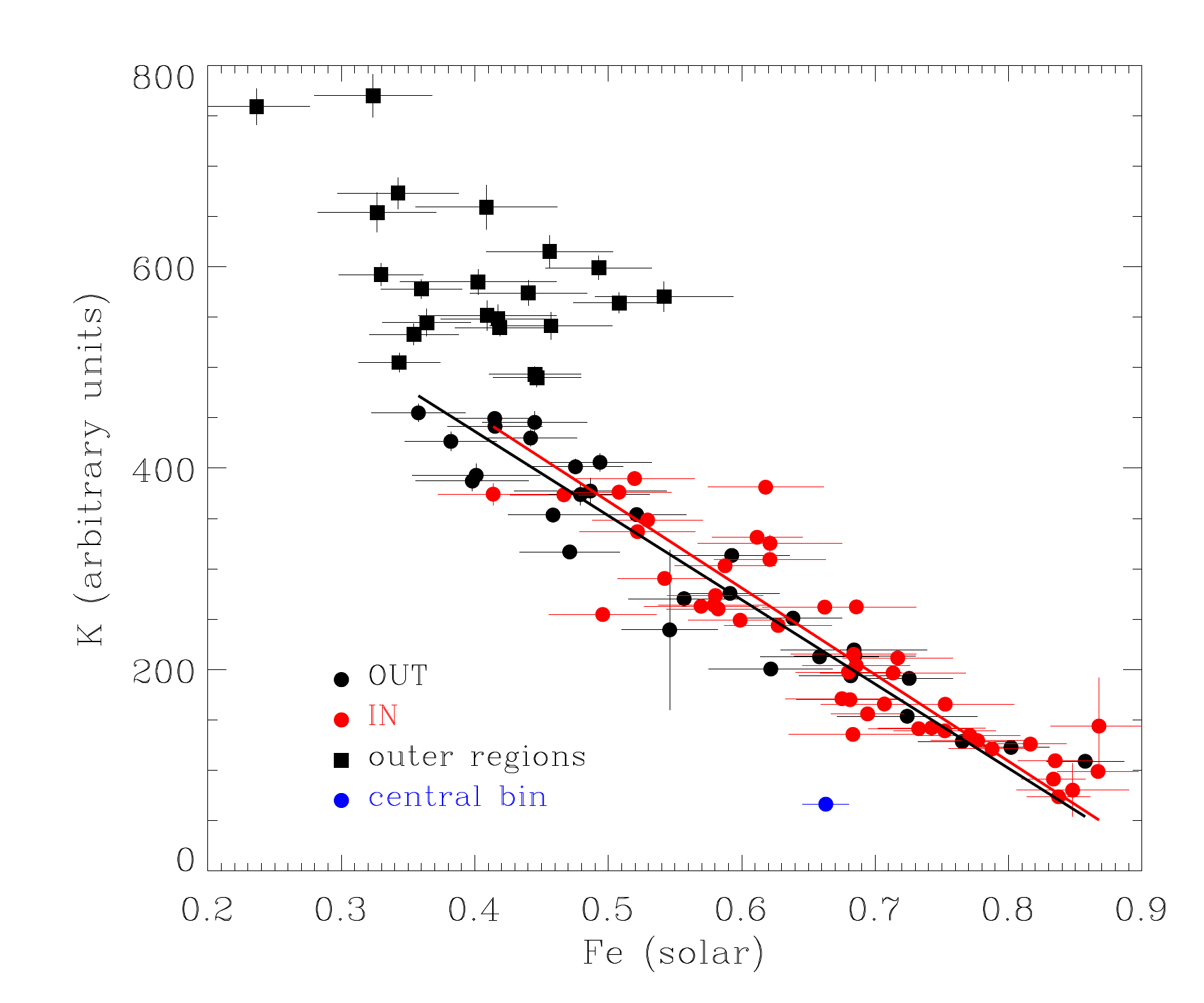} 
\caption{Entropy - Fe abundance correlation. Red marks the IN polygons,
black marks the OUT polygons. The red and the black lines are the best fits for IN and OUT points 
respectively.
The best fits have been derived using the BCES estimator after excluding the outermost regions 
(black squares) and the spectrally complex central bin (blue point).
}
\label{fig:bces_fit}
\end{figure}

Fig. \ref{fig:bces_fit} shows the well-known \citep{Degrandi:2004,Leccardi:2010}
correlation between the metallicity and the gas entropy (K-Fe hereafter). 
We exclude from our analysis the outlying blue point which 
corresponds to the spectrally complex central bin (see Appendix \ref{sec:spec_models} for details).
Outer regions (filled black squares in figure) deviate from the correlation: at those distances the entropy 
continue to increase while the metallicity reaches a constant value
($ \sim 0.3-0.4$ \zsun). 
The remarkable result is that the regions IN the spiral appear to follow the same relation as the regions OUT 
of the spiral (red and black points respectively). 
We have quantified this qualitative result with the BCES estimator fitting routine (which allows best fit determination for measurements with 
errors in both variables; \citealp{AB_BCES:96}).
Fitting with $K= K_0 + r Fe$ we find $K_0=798 \pm 45$ and $r=-861 \pm 65$ for IN regions and 
$K_0=771 \pm 29$ and $r=-836 \pm 51$ for OUT regions. We point out that outer regions, were the metal abundance levels off, have been 
excluded from the fit, best fits are also shown in Fig. \ref{fig:bces_fit}.

In light of the striking similarity of the K-Fe relation for IN and OUT regions,
we have verified if it may be spurious in nature. More specifically we have checked that the observed IN K-Fe relation is not due to the 
neglection of multi-phase structure in our spectral analysis. 
The interested reader is referred to Appendix \ref{sec:fekcheck} for a detailed description of this verification.
The fact that the sloshing mechanism does not destroy the correlation between metal 
abundance and entropy,
has far reaching consequences which will be discussed
in detail in the next section.

\subsection{Thermodynamical processes at work in the ICM}
\label{sec:thproc}
During the sloshing process, mechanisms that might erase the density and temperature
differences between the different phases of the plasma that come into close contact
must be suppressed. 

The main processes able to erase temperature differences on short timescales are conduction and 
convection.
The last process operates on the gas by physically mixing the two different components, while
thermal conduction 
brings the gas to the thermal equilibrium through heat exchange without mixing the gas particles.
Thermal conduction is the most efficient mechanism to erase temperature differences.
In the Spitzer \citep{Spitzer} regime, thermal conduction operates on a timescale of  
$$t_S \sim k{\ell}^2 n_e/\kappa_S \sim  1.3 \times 10^{-2} n_{e,2} {\ell}_{100}^2 ~(T_{10})^{-5/2}~~{\rm Gyr}
 $$ \citep{MM_A754:2003,Rasmussen:2006} where $k$ is the Boltzmann constant, $\kappa_S$ is the Spitzer heat conductivity, $T_{10}$ is the temperature 
in units of 10 keV, ${\ell}_{100}$ the region width ${\ell}$ in units of 100 kpc, $n_{e,2}$ the electron density $n_e$ in units of $2\times 10^{-3}$ cm$^{-3}$. 
In our case, ${\ell}$ can vary from few kpc (the typical width of a cold front if we want to estimate the timescale 
to erase discontinuities) to some tens of kpc if we want to bring the whole spiral in thermal equilibrium. 
The spiral reaches the maximum width in the tail, with a radius of 
$ \sim 60$ kpc. The electron density is about 
$n_e \sim 3\times 10^{-3}$cm$^{-3}$ at the NNW cold front, while the southern most external cold front at $150$ kpc
has a density of $n_e \sim 1.5\times 10^{-3}$cm$^{-3}$.
To derive a rough estimation of $t_S$ for the IN gas, we assume  $n_e = 2\times 10^{-3}$cm$^{-3}$, 
and  $T = 4.5$ keV and ${\ell}_{100}$ spanning the range $[0.06-0.6]$.
In this hypothesis, the cold fronts would be erased  
within $t_S = 3.3\times 10^{-4}$ Gyr, (assuming ${\ell}_{100} = 0.06$) and the gas in the spiral would reach the thermal equilibrium with the environment 
within $t_S = 3.3\times 10^{-2}$ Gyr (${\ell}_{100}=0.6$). 
As pointed out in Sect. \ref{sec:age}, the lifetime of the cold fronts and of the spiral 
structure in A496 is about $\sim 0.6$ Gyr, 
three orders of magnitude higher than the time required to erase the cold front features and one order of magnitude
higher than the time necessary to reach the thermal equilibrium and erase the spiral structure:
the spiral would be quickly destroyed if conduction were not suppressed along all the spiral edge.
While suppression of conduction at cold fronts has been reported by several authors 
\citep[e.g.][]{Ettori_conduc:2000,Vikhlinin1:2001} this is, to the best of our knowledge, 
the first time that suppression is reported between gas in and out of the spiral. 
Magnetic fields are likely to be responsible for such suppression.
Magnetic fields with strengths of the order of $\mu$G are commonly observed in galaxy clusters 
\citep[for recent reviews]{CT:2002,Ferrari_rev2008}. Radio mini-halos are also observed in cool 
cores of 
sloshing clusters \citep{Govoni:2009,Giacintucci:2011, Giaci:2014} suggesting a possible association
between 
the radio emission and sloshing cold fronts \citep{MG:2008,ZuHone:2011,ZMBG:2013}. Simulations 
\citep{Asai:2005, Asai:2007, Lyutikov:2006,DP:2008}
show that sloshing can amplify the magnetic field strength up to an order of 
magnitude and that magnetic fields drape around 
the front surfaces inhibiting the conduction and the heat exchange across the front.

Various mechanisms  can, potentially, operate on the entropy and metal abundance of the sloshing gas. 
Conduction, if it were to occur at the Spitzer level,  would rapidly transfer heat from the hotter 
environment to the cooler sloshing gas. 
Indeed our calculations show that the typical timescale for heat transfer over scales of few tens of kpc is significantly 
shorter than the dynamical timescales associated to the sloshing process. 
In other words, the motions are sufficiently slow that, as the  sloshing gas moves back and forth, 
it should suffers significant heating from the ambient ICM. Effective heating would therefore move 
gas in the spiral away from the 
K-Fe correlation defined by gas outside the spiral. Unless, 
of course,  
heating were accompanied by dilution of the IN gas with OUT gas at just the right rate to maintain
the IN points on the K-Fe correlation defined by OUT points. 
Assuming that conduction is significantly suppressed, mixing of IN and OUT gas could both heat 
the IN gas and dilute its metal content. 
However, it would have to operate on spatial scales we cannot resolve (smaller than a few kpc) and, most importantly,  the mixing 
would have to to occur at just the right rate 
as the IN gas must move along the K-Fe correlation defined by the OUT gas.
Moreover: 1) the same processes inhibiting heat transfer by more than one order of magnitude would likely suppress mixing and 
2) as already pointed out in Sec. \ref{sec:fe-slosh}, continuous mixing of the IN with OUT gas shuold lead to smaller difference between IN and OUT  
metallicity at larger radii, where the IN gas would have had more time to mix with the OUT gas.
In other words, we would need a complex mechanism that
is fine tuned to keep the points on the K-Fe correlation.
It must also be noted that this mechanism should at the same time preserve the discontinuities we observe at the cold fronts.
The alternative, namely that regions on the spiral lie along the relation defined by regions
outside the  
spiral because both conduction and mixing mechanisms are heavily suppressed, seems to provide
a much simpler 
and natural explanation.

Some MHD simulations \citep[e.g.][]{ZuHone:2013} show that some anisotropic conduction 
can be present in the ICM allowing 
heat exchange and entropy increase for the cool gas, preserving the characteristic sharp 
(though reduced) jumps. However, if such a mechanism is indeed operating, it must be rather
ineffective, otherwise it would end up removing the IN 
regions from the K-Fe relation defined by OUT regions.

\subsection{Sloshing and heating}

Sloshing may contribute to the heating of the cooling core. 
A possible way is through turbulence: sloshing gas motions may induce turbulence and turbulent energy 
may be dissipated into heat. However, simulations  show that turbulent energy in sloshing cores is 
a negligible fraction of the thermal energy \citep{Vazza_turb:2012,ZMBG:2013} and also 
turbulent diffusion is low \citep{Vazza_turb:2012}. While this mechanism could be responsible 
for the presence of radio-mini halos  in sloshing cores they likely provide little contribution to the core heating.
An alternative way to quench the cooling in a sloshing picture is through 
mechanical energy transferred to the ambient ICM via $pdV$ work.
During the sloshing, the central cool gas is progressively lifted upwards and comes
into contact with the hotter 
environment, which is less dense and with a lower pressure. Gradually, the cooler and denser gas will
expand adiabatically to achieve 
pressure equilibrium with the surrounds. While expanding the gas does $pdV$ work on the
surrounding atmosphere.
We make use of our observation to test whether the expanding gas con provide sufficient $pdV$ work to offset 
radiative cooling. We estimate the mechanical power exerted by the sloshing gas and transferred on the environment 
by following the evolution of the gas currently residing in the tail of the spiral.
Under the assumption that the sloshing process is close to adiabatic, this gas must have originally 
resided some 60 kpc from the center.
By deprojecting and assuming entropy conservation we derived that the work done by the 
gas rising from $~60$ kpc to $~150$ kpc, is about $W = \int {pdV} = 8.08 
\times 10^{58}$ erg. The estimated timescale for the mechanism is $t \sim 0.7$ Gyr providing a 
power $W/t \sim 3.66 \times
10^{42}$erg/s. Using the deprojection  algorithm \citep{Ghizzardi_M87:2004,Ettori_depro:2002a, Ettori_depro:2002b},
we also evaluated the bolometric 
luminosity within the cooling radius ($\sim 75$ kpc) and we obtained $L_{cool}= 5.16 \times 10^{43}$ erg/s, 
which is more than one order of 
magnitude larger than the mechanical power supplied by the expanding gas. The cooling cannot 
be quenched by 
the sloshing mechanism. In addition, if the sloshing gas and the ambient gas do not mix significantly, 
as our results suggest,
the cool gas will eventually sink back 
to the bottom of the potential well. During this reverse process the low entropy gas will 
be compressed by the ambient medium and, in the 
adiabatic limit, the same amount of $pdV$ work originally done by this gas will now be exerted on it.
Net heating will be small, as it will be related to 
deviations from the adiabatic evolution of the sloshing gas,
which, as discussed in Sect. \ref{sec:thproc}, must be small.

\subsection{South-east metal excess}
\label{sec:SE_excess}

The metallicity map (see Fig. \ref{fig:Z_Zres_midi}) shows a hint of a metallicity 
excess in the south-east 
(see also \citealp{Ghizzardi:13p}).

\begin{figure}
 
\centering
\hspace{-1.1 truecm}
\includegraphics[angle=0,width=9.5 truecm]{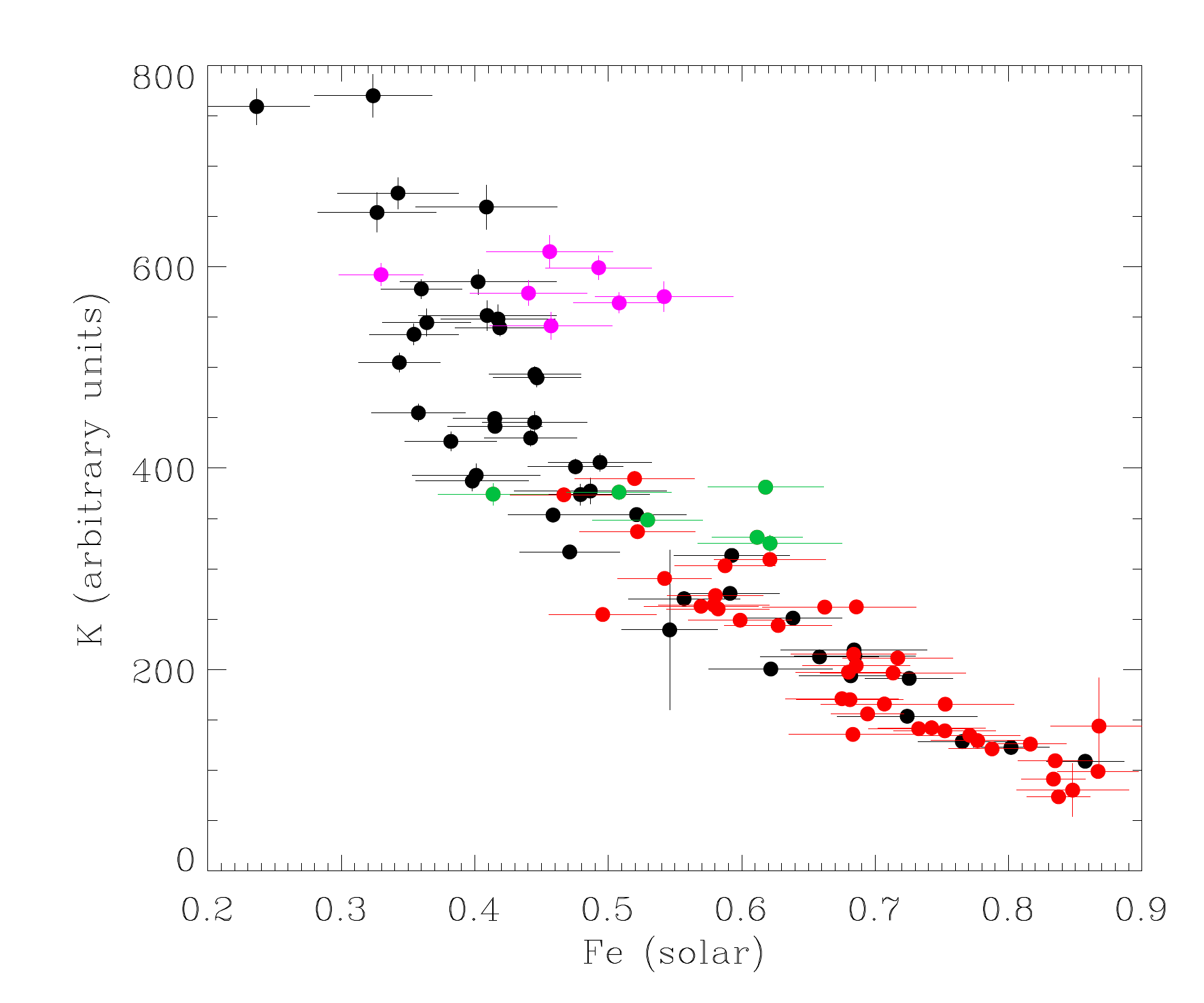} 
\caption{K-Fe correlation. Red marks the IN polygons,
black marks the OUT polygons, magenta marks SE polygons OUT of the spiral, and green 
marks SE polygons IN the spiral. 
All the SE regions show a metal excess and introduce a significant scatter in the  plot.}
\label{fig:scatter}
\end{figure}

In Fig. \ref{fig:scatter}, we marked, as magenta circles, the south-east 
regions located outside
the spiral.
The plot confirms that these regions have a higher metal content, significantly higher than 
that of the regions 
having a comparable entropy level.
The nature of this excess is not of immediate interpretation. The excess is
located just outside the
spiral pattern. This could suggest that some metals have been injected in the surrounding ICM. 
These regions are also close 
to the regions where Kelvin-Helmoltz instabilities might be at work (Sect. \ref{sec:KH}), 
facilitating the mixing of gas.
However, we note that the south-east regions located on the spiral show a similar metal excess. These 
regions are marked with green 
circles in Fig. \ref{fig:scatter}. It is remarkable that all these regions (magenta and green points) 
introduce a significant scatter in the K-Fe plot. Indeed, if we remove these points from the plot, 
the K-Fe correlation is significantly tighter. 
Although we lack a deep understanding of the origin of this metal excess the fact that it involves both 
regions IN and OUT of the spiral leads us to think that it is likely not due to mixing or 
diffusion processes.
This is also supported by the fact that all the other regions of the cluster lie on the K-Fe 
correlation. 
A more attractive alternative is that this excess might trace some cluster asymmetry predating
the onset of the 
sloshing.

\section{Summary}
\label{sec:summary}
We analyzed two long XMM-Newton observations ($\sim$ 120 ksec) of A496. Taking advantage of the XMM-Newton large collecting
area and spectral resolution the quality of the data allowed a detailed characterization of the metal abundance for this cluster.
Our main results concern the detection of sloshing cold fronts and the connection between the metal distribution and the sloshing.

\begin{itemize}
 \item {We detect 3 and characterize 2 of the 4 cold fronts hosted in A496.
  The main cold front is located in the NNW direction at a distance $\sim 60$ kpc from the X-ray peak. We also detect two cold fronts 
  in the south direction $\sim 35$ and $\sim 160$ kpc from the center. 
 }
 \item{We detect the low temperature/entropy spiral feature found in many sloshing cores. 
 All the cold fronts are located along the edge of the spiral; the southern outermost cold front is placed at the tail of the spiral.}
 \item{The Fe abundance drops abruptly across cold fronts. A similar but less significant trend is seen in Si and S.}
 \item{Our metallicity map shows that plasma in the spiral has a higher metal content than the surrounding medium.}
 \item{We find that regions in the spiral follow the same K-Fe relation as regions outside the spiral.}
 \end{itemize}
 
 All these results allow us to build a fairly articulated picture of the evolution of the ICM in the core of A496:

 \begin{itemize}
\item{Using fiducial numbers for the age of the sloshing structure we find that conduction between the gas in the spiral and the ambient medium must be suppressed by more 
than one order of magnitude with respect to Spitzer conductivity.}
\item{Under the reasonable assumption that prior to the onset of sloshing the ICM was stratified according to entropy, we infer that while the 
low entropy metal rich plasma is uplifted through the cluster atmosphere it suffers little or no heating/mixing with the ambient medium.}
\item{While sloshing appears to be capable of uplifting significant amounts of gas, the limited heat exchange and mixing between gas in and outside the spiral implies 
that this mechanism is not at all effective in:
1) permanently redistributing metals within the core region (the cooler metal richer gas will eventually fall back to the center) and 2) heating up the coolest and densest gas, thereby providing little or no contribution to staving off 
catastrophic cooling in cool cores.}
\item{The only indication we have that some mixing might (and we stress the might) be occurring comes from regions E and W of the center, 
where the spiral edges appear to be smeared out.}
\item{Finally, within the ``quasi-adiabatic sloshing scenario'' we have outlined,  the excess we  
observe both in and out of the spiral SE of the center, most likely traces some cluster asymmetry predating the onset of sloshing.}
\end{itemize}

% \begin{acknowledgements}
%      ..'azie
% \end{acknowledgements}

\bibliography{biblio}
\bibliographystyle{aa}
%\begin{thebibliography}{}

%\end{thebibliography}
\appendix
\section{Spectral models comparison.}
\label{sec:spec_models}

We have performed a detailed analysis of the temperature structure
of the regions of interest in A496 (annular sectors and polygons), 
searching for possible multi temperatures components in the ICM 
spectra.

In this Appendix we present the results from the analysis of
each spectrum with the two different spectral models described in
Sect. \ref{sec:spec_analysis}: the one temperature {\it vapec} model (1T model), and 
the multi-temperature GDEM model \citep{Buote:2003}.

\begin{figure}\centering
   \hspace{-0.2 truecm}
   \includegraphics[width=9.5cm]{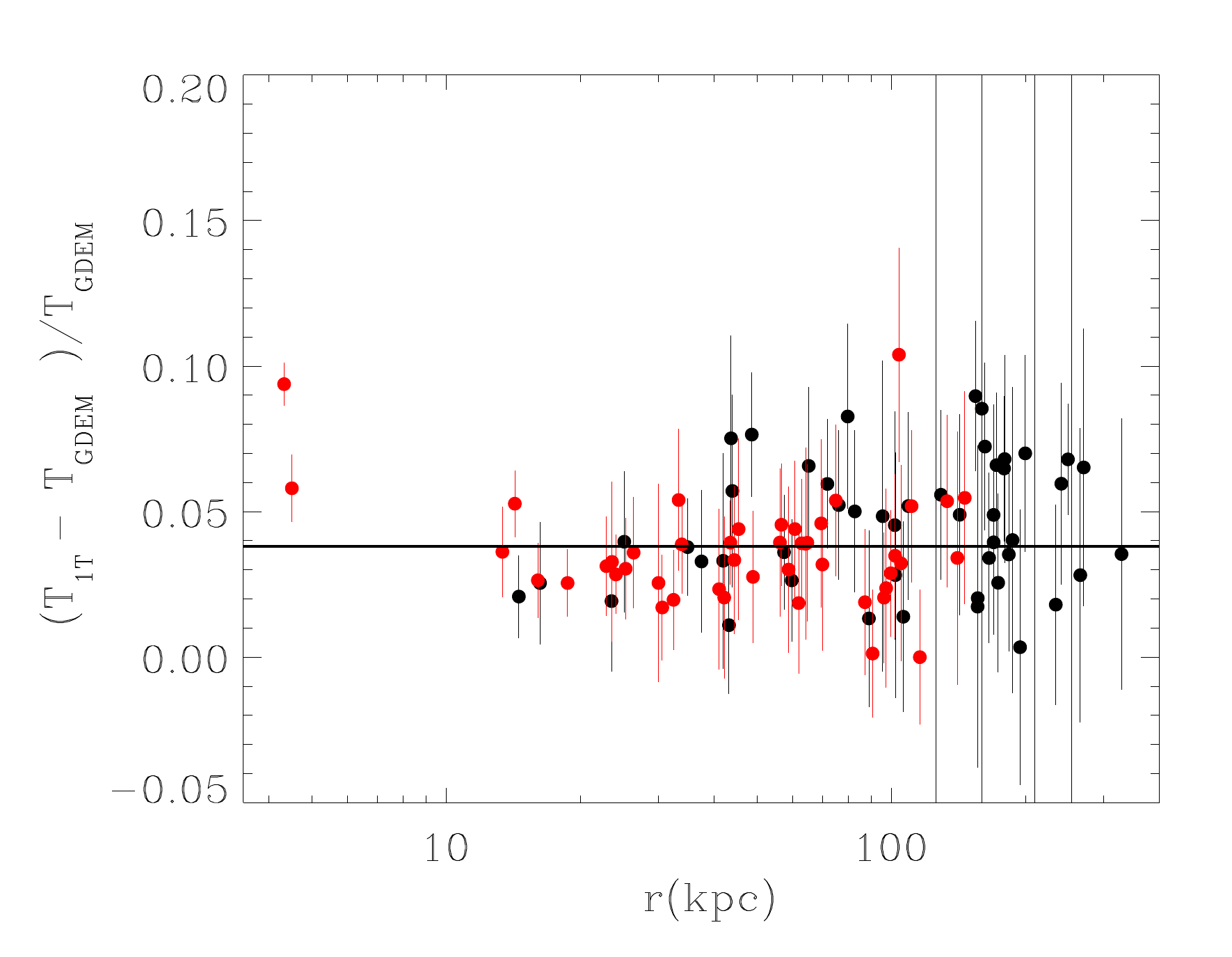}
      \caption{Relative temperature differences between the
      1T and GDEM models as a function of the distance from the
      center of the cluster. Red marks the IN polygons, black 
      marks the OUT polygons.  }
         \label{fig:temp_diff}
   \end{figure}

In Fig. \ref{fig:temp_diff} we show the relative differences between the temperatures
estimated with the 1T model, $T_{1T}$, and the GDEM model, 
$T_{\rm GDEM}$, i.e. $(T_{1T}-T_{\rm GDEM})/T_{\rm GDEM}$, plotted as a 
function of the physical distance from the cluster center of the regions, 
for the 98 polygons used to extract the spectra (see Sect. \ref{sec:spiral} and Fig. \ref{fig:spiral_in_out}).
As is the main text red circles mark the polygons in the spiral (IN) and
black circles mark the polygons outside the spiral pattern (OUT).
In the Fig. \ref{fig:temp_diff} the central bin is the only one that shows a temperature 
difference significantly larger than $\sim 5\%$, whereas in all other bins 
the differences are around $3-6\%$ and consistent with this value within
$1-2\sigma$. 
The best fitting constant model, with the exclusion of the first bin, gives 
$0.038\pm 0.002$ and is plotted in Fig. \ref{fig:temp_diff} as a solid line.

\begin{figure}\centering
   \hspace{-0.2truecm}
   \includegraphics[width=9.5cm]{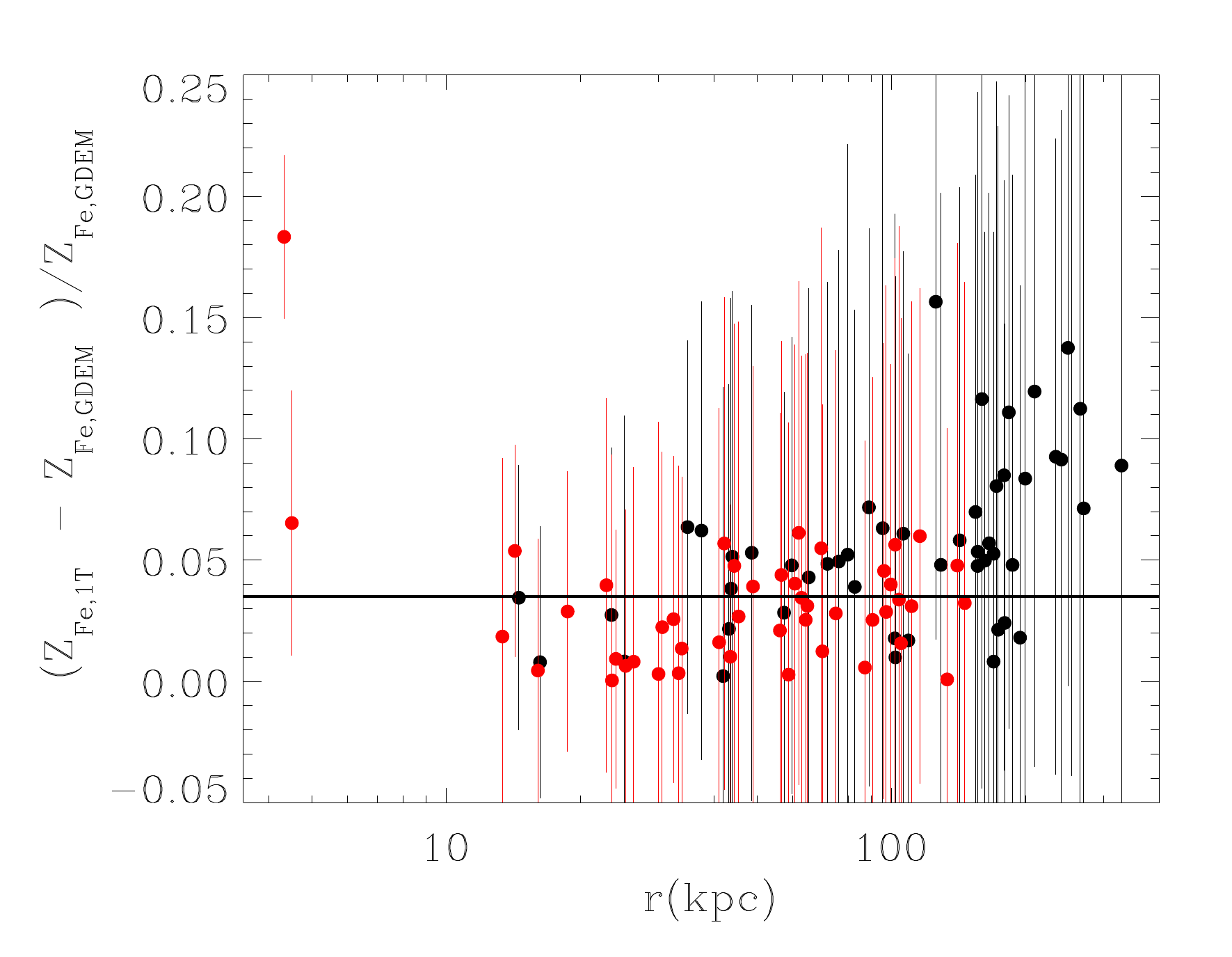}
      \caption{ Relative Fe abundance differences between the
      1T and GDEM models as a function of the distance from the
      center of the cluster. Red marks the IN polygons, black 
      marks the OUT polygons.  }
         \label{fig:fe_diff}
   \end{figure}

The relative Fe abundance differences, 
$(Z_{\rm Fe,1T}-Z_{\rm Fe,GDEM})/Z_{\rm Fe,GDEM}$, 
give similar results (see Fig. \ref{fig:fe_diff}). In this case, the best fitting 
constant model gives $0.035\pm0.01$. Symbols and line have the 
same meaning as in Fig. \ref{fig:temp_diff}.

We can ascribe the small $\sim 3\%$ bias towards slightly larger temperatures
and Fe abundances from the 1T model to well known calibration problems in the 
EPIC detector between the soft (0.7-2 keV) and hard (2.-8. keV) X-ray bands 
\citep[e.g.][]{Nevalainen:2010}.

From the negligible differences between temperatures and Fe abundances
in the two spectral models it is clear that, if present, a multi-temperature 
component in the ICM provides only a modest contribution to the
total emission. 
More interestingly for the purposes of our work,
the value of the measured Fe abundance is essentially unaffected by 
a possible multi-temperature structure.

We exclude from our analysis the central bin, which is the only one where
the relative differences are significative in both cases. 
This central cluster region could
be contaminated by the presence of the central AGN or by a true 
multi-phaseness of the ICM.  

%----------------------------------------------------------- 
%----------------------------------------------------------- 

%----------------------------------------------------------- 
%----------------------------------------------------------- 

\section{Pressure jump across the cold fronts}
\label{sec:app_press}

In Sect. \ref{sec:prat} we derived the pressure jump across the main cold fronts: 
NNW, divided into two sectors 30$^{\circ}$-75$^{\circ}$ and 75$^{\circ}$-120$^{\circ}$
to account for its boxy morphology, and S2. 

To a first approximation, we derived the pressure jump across the front by using the electron 
density and the temperature 
in the two bins just inside and outside each edge. 
Namely, we model the electron density profile 
with a broken power law:

\begin{equation}
n=\left\lbrace
\begin{array}{ll}
  n_{in} \left( {r \over r_{CF} }\right)^{-\alpha_{in}} & r < r_{CF} \\
  n_{out} \left( {r \over r_{CF}} \right)^{-\alpha_{out}}  & r > r_{CF} 
\end{array}
\right.
\end{equation}

where $n_{in}$ and $n_{out}$ are the electron densities at the cold front position $r_{CF}$, 
on the inner and outer side 
of the edge respectively. 
We projected the emissivity along the line of sight and  we fit the surface brightness profiles 
of each sector of interest 
to derive the values of $n_{in}$, $n_{out}$, $\alpha_{in}$, and $\alpha_{out}$.
The fitting process has been divided into two steps: we initially fit only the outer part of
the profile 
(outside the cold front) 
where only the external component is present and derived the $n_{out}$ and $\alpha_{out}$ parameters.
Successively, we 
fixed the external component and fit the whole profile to derive the inner component parameters.
Finally we derived the pressure jump $p_{in}/p_{out} = n_{in}T_{in}/n_{out}T_{out}$ 
where the temperature values $T_{in}$ and $T_{out}$ in the two bins close to the front position 
have been obtained through the spectral analysis (see Fig. \ref{fig:profili_z}).
The pressure jumps obtained through this method are reported in the second column (approx pressure jump)
of Table \ref{tab:pj}.

\begin{figure*}
  % \centering
   \includegraphics[angle=0,width=6. truecm]{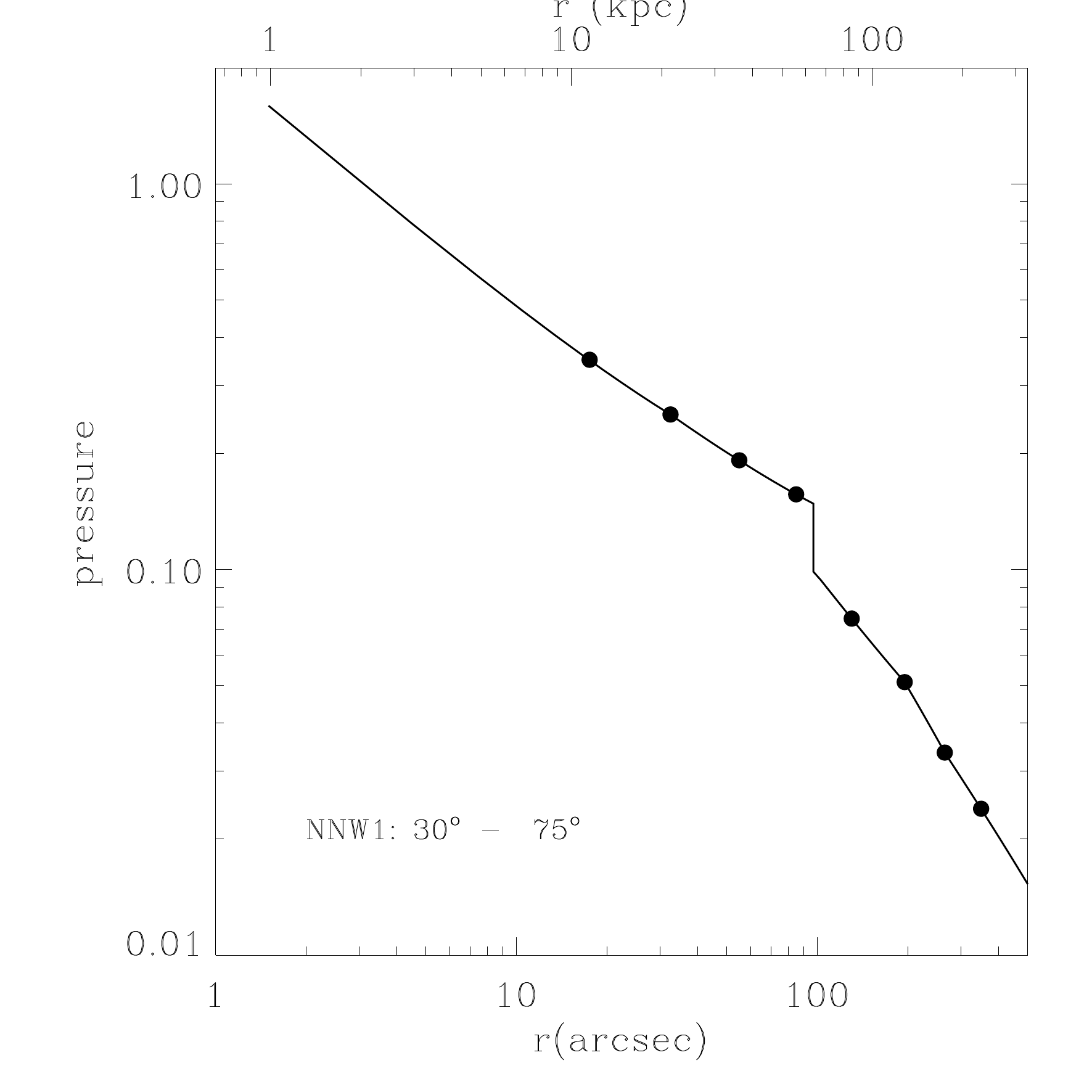} %6 cm in non -referee format
   \includegraphics[angle=0,width=6.truecm]{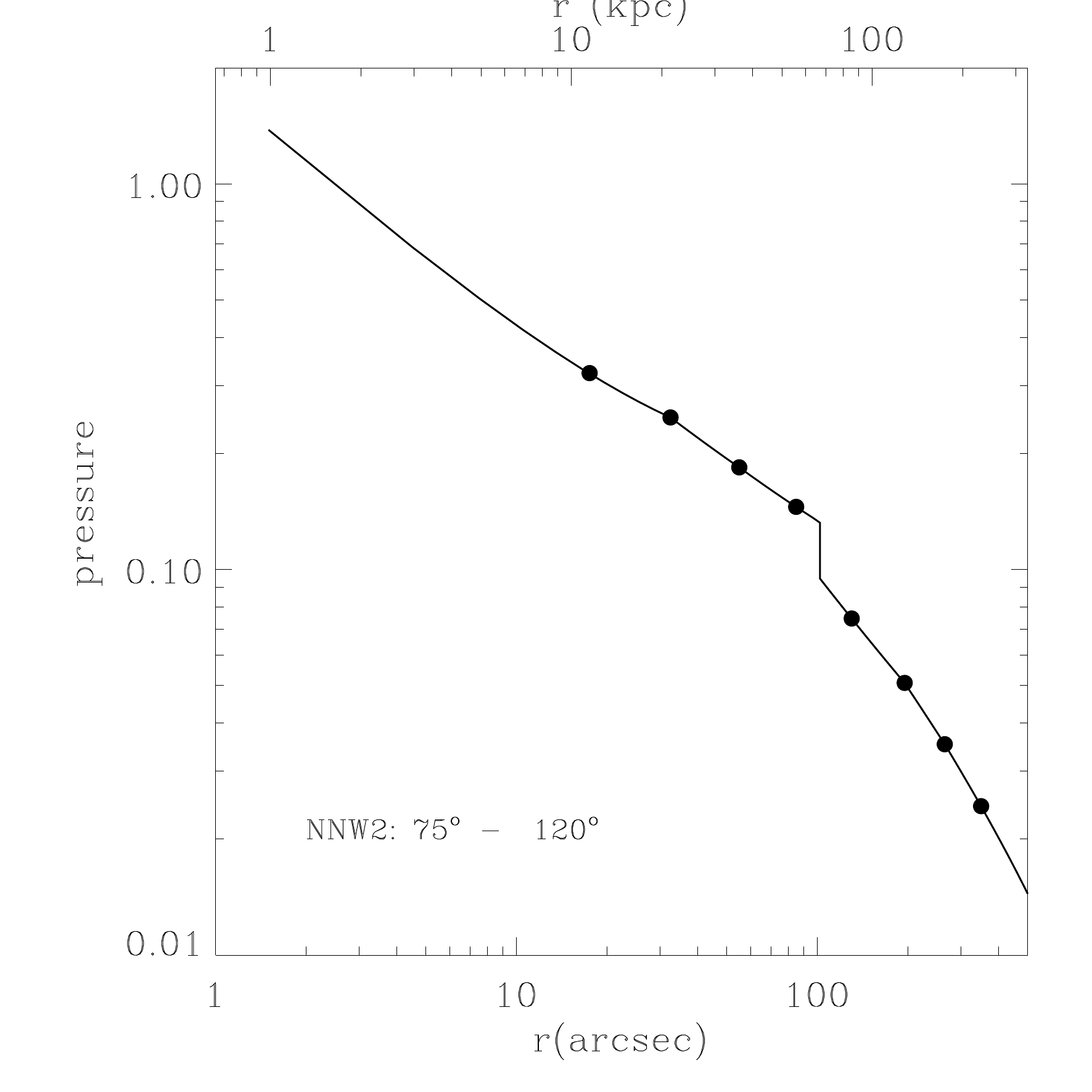}
   \includegraphics[angle=0,width=6. truecm]{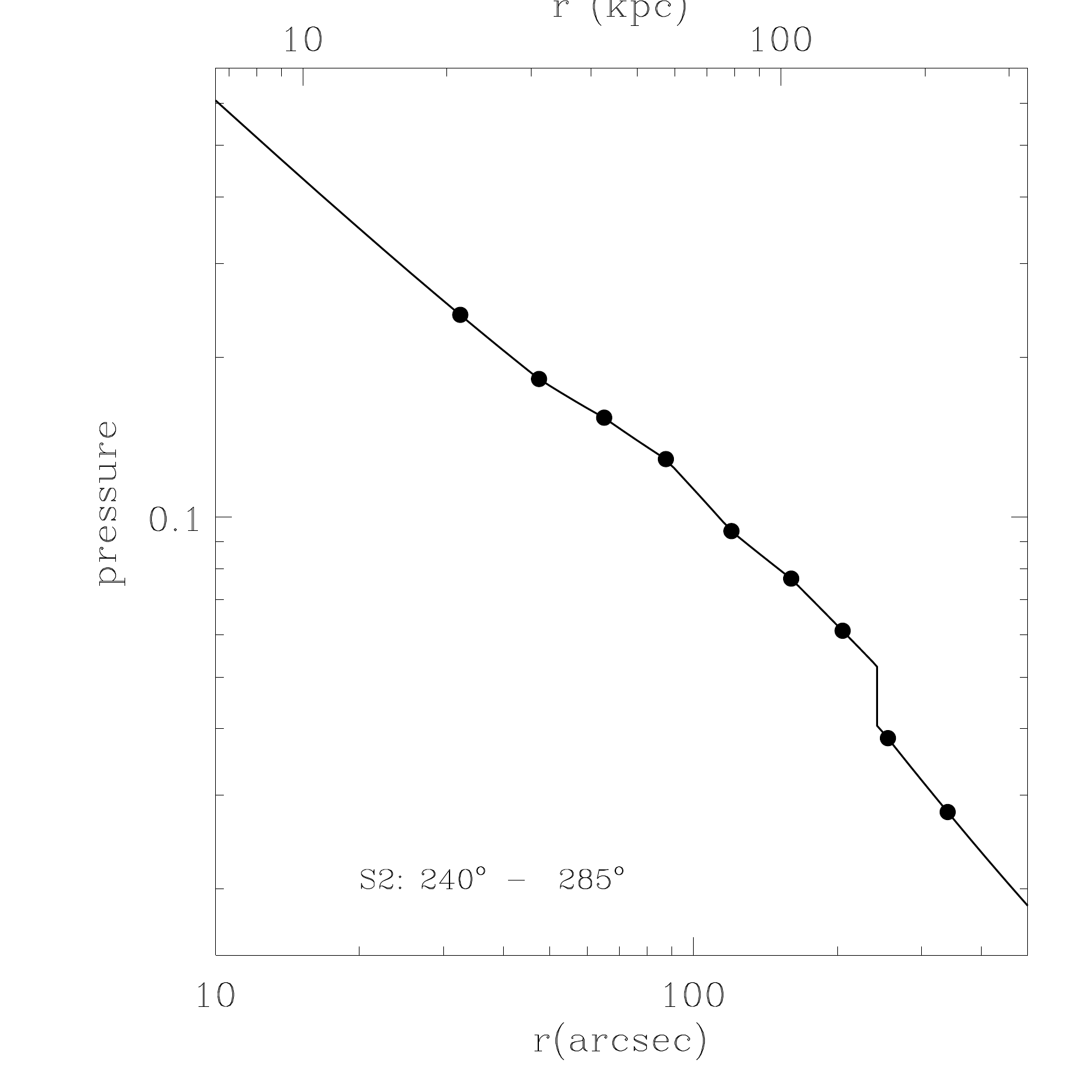}
   \caption{Pressure (arbitrary units) profiles obtained with the two different methods (see text) in the three cold fronts sectors $30^{\circ} - 75^{\circ}$ (left panel), 
$75^{\circ} - 120^{\circ}$ (middle panel) and $240^{\circ} - 285^{\circ}$ (right panel). Solid line is the pressure profile obtained 
by using the interpolated temperature profile, filled black circles correspond to the pressure obtained by using the original $T$ profile.  } 

\label{fig:press_stagn}%
    
\end{figure*}

A more accurate estimate for the pressure profile can be obtained by multiplying the electron 
density profile with the temperature profile. 
However, the temperature profiles (see Fig. \ref{fig:profili_z}) are evaluated on bins which are 
significantly larger than 
the bins available for surface brightness (and electron density)  profiles.
We then interpolated the temperature profiles to reconstruct a temperature profile on the same bins 
used for the density profile.
We performed two independent interpolations for the points inside and outside the cold front. 
For the outer 
regions we excluded the points close to the cold front in order to keep far away from the 
stagnation point, 
profiles in this region are computed by 
extrapolating measures at larger radii. 
We note that since bins used for the temperature profiles are large to allow reliable spectra fitting and the number of bins 
is quite small, we did not deproject temperature.

\begin{table}
\caption{Pressure jumps for the three cold front sectors}
\label{tab:pj}
\centering
\begin{tabular}{l c c}
\hline
sector  & approx pressure jump & pressure jump \\
\hline\hline
NNW: $30^{\circ} - 75^{\circ}$ & $1.40 \pm 0.03$ & $1.5 \pm 0.05$  \\
NNW: $75^{\circ} - 120^{\circ}$ & $1.33 \pm 0.02$ & $1.40 \pm 0.05$ \\
S2: $240^{\circ} - 285^{\circ}$ & $1.28 \pm 0.03$  & $1.29 \pm 0.05$ \\
\hline
\end{tabular}

\end{table}

In Fig. \ref{fig:press_stagn} we plot the pressure profiles (solid line) derived 
for the three cold fronts sectors. 
The corresponding pressure jumps are reported in the last column of Table \ref{tab:pj}.
As expected, the jumps measured from the pressure profiles are slightly larger than the approximated 
ones: $\sim 7 \%$ and $\sim 5 \%$ for the two NNW sectors 
and ~0.7 \% for S2.

In Fig. \ref{fig:press_stagn} we also overplotted (black circles) the $nT$ profiles obtained from the
original
temperature profiles. 
The figure confirms that the two profiles match and that the differences between the results obtained 
with the two 
procedures are small.
 
\section{Testing the robustness of the K-Fe correlation}
\label{sec:fekcheck}

In this appendix we address the robustness of the correlation (K-Fe) 
between entropy and metal abundance. We want to investigate whether the match between 
IN and OUT regions in the K-Fe plot might be due to the single temperature model employed to fit 
the spectra.
Namely, if, in the regions on the spiral, there is some gas mixing,  two different gas phases would coexist 
and the 
1T fitting might provide an averaged value for $T$, $Z$, $n_e$ which could preserve the correlation even 
if the gas is made up
of two distinct components. Moreover, even if the gas on the spiral is not affected by mixing, 
projection effects can lead 
to the same bias since the analyzed spectrum would contain the two components. We fitted the spectra
with a 
two temperature model, but using this model does not improve
the quality of the fit.
We therefore took a different approach, namely we built a composite spectrum containing the
spectra of two different regions lying at opposite sides in the K-Fe plot.
This simulates both a region containing mixed gas and a region affected by projection.

\begin{figure}
 
\centering
\hspace{-1.1 truecm}
\includegraphics[angle=0,width=9.5 truecm]{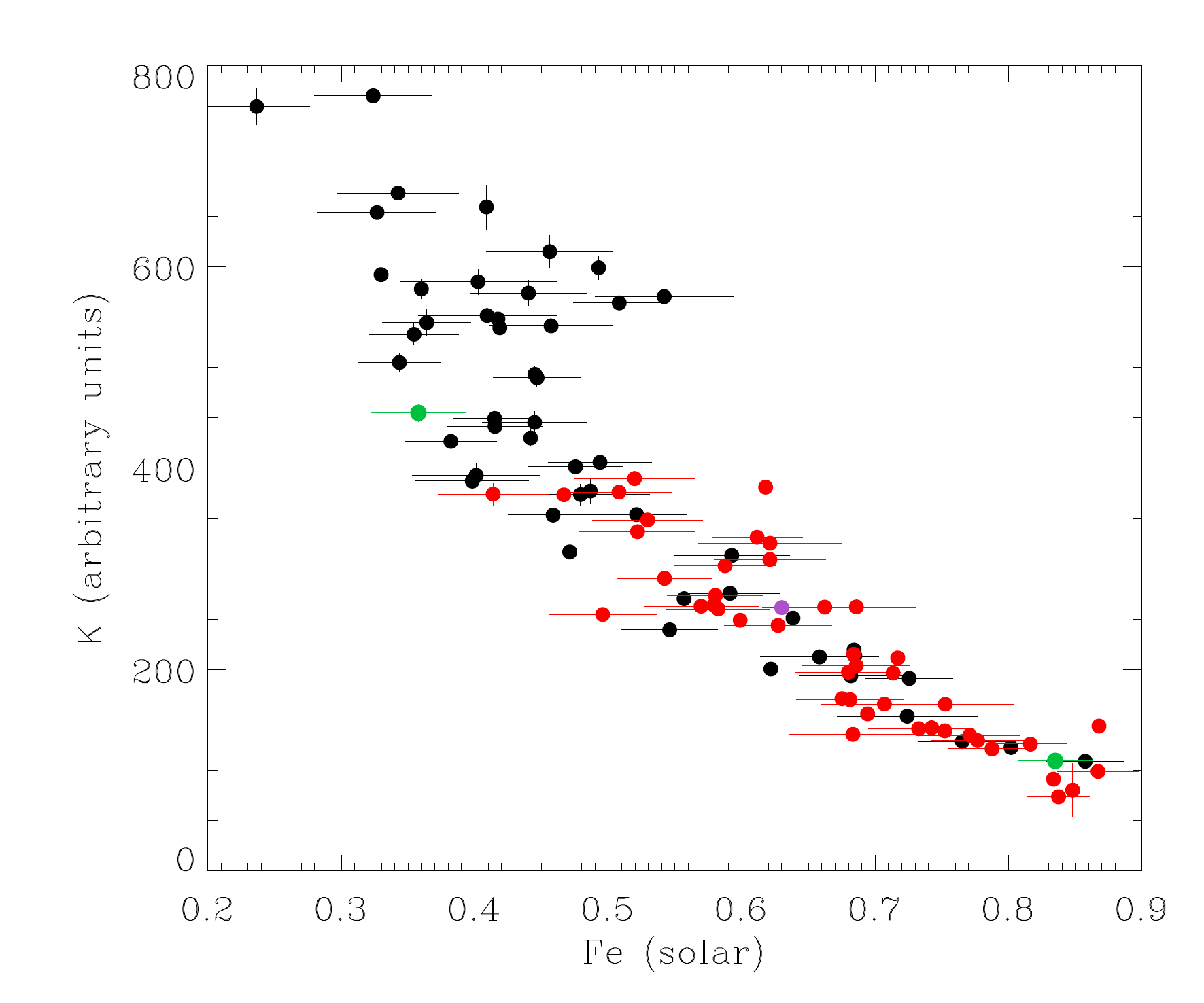} 
\caption{K-Fe correlation. Red marks the IN polygons,
black marks the OUT polygons. Green points mark regions $\# 2$ and $\#71$ used for the test 
and the purple point represents
the composite region $\# 2+ \# 71$.}
\label{fig:mixing}
\end{figure}

We chose regions $\#  2$ and $\# 71$ (green points in Fig. \ref{fig:mixing}).
The lowest temperature region ($\# 2$) is on the spiral feature at the center 
of the cluster, 
while region $\# 71$ is out of the spiral in the northside $\sim 200$\as ($\sim 120 $ kpc) 
from the center.

\begin{table*}
\renewcommand{\arraystretch}{1.5}
 \caption{Best fit values for temperature, and  metal abundance % and normalization 
 obtained using 1T and 2T model for regions $\# 2$, $\# 71$ and for the composite spectrum $\# 2+ \# 71$. The two last columns report the F-test value and the 
 corresponding probability.}
\label{tab:mixing}
\centering
\begin{tabular}{l |c c c | c c c c c|c|c}
\hline
&\multicolumn{3}{c|}{1T}& \multicolumn{5}{c|}{2T} & F & P \\

%\vspace{0.2 truecm}
& $T$ (keV) & $Z_{Fe}$ (solar) & $\chi^2/dof$ &  $T_1$ (keV) & $Z_{1,Fe}$ (solar) &  $T_2$ (keV) & $Z_{2,Fe}$ (solar) & $\chi^2/dof$ & & \\  
\hline
$\# 2$ & $2.58 \pm 0.02$  &  $0.83 \pm 0.04$  & 576.2/455 & $1.72^{+0.24}_{-0.05}$ & $ {\rm unconstrained}$  & $2.89 \pm 0.4$ &  $0.57^{+0.06}_{-0.11}$ & 543.0/452 & 9.21 &  $6.33 \times 10^{-6}$ \\
$\# 71$ & $4.80 \pm 0.10$  &  $0.34 \pm 0.04$  & 520.5/540 & $2.39^{+1.03}_{-0.43}$ & $0.13^{+0.10}_{-0.06}$ & $6.85^{+2.14}_{-0.35}$ & $0.48^{+0.32}_{-0.09}$ &  508.9/537 & $4.09$ & $6.93 \times 10^{-3}$   \\
%\vspace{0.2 truecm}
$\#2 + \# 71$ & $3.29 \pm 0.03 $  &  $0.61 \pm 0.01$  & 760.3/676 & $2.27 \pm  0.13$ & $ 0.55 \pm 0.05$ & $ 6.84 \pm 0.7$&  $0.49 \pm  0.07$ & 652.5/673 & 63.27 & $4.76 \times 10^{-36}$ \\

\hline
\end{tabular}

\end{table*}

We checked the count rates of both regions $\# 2$ and $\#71$ and verified that they have similar values, so that the two regions contribute at the same level to the total 
spectrum. 
This mimics a region where mixing occurs with comparable filling factors for the two different gas phases.
We first fit the composite spectrum with a 1T model finding, not surprisingly, best fit values of the temperature and of the metal abundance (see Table \ref{tab:mixing}) 
that are in between those measured for regions $\# 2$ and $\# 71$. This also leads to a point in the K-Fe plot that is indeed still on the relation (see Fig. \ref{fig:mixing}), 
however, in this case, the fitting procedure returns a much larger $\chi^2$, $760.3$ for $676$ dof against $\chi^2 = 576.2$ for $455$ dof for region $\# 2$ and
$\chi^2 = 520.5$ for $540$ dof 
for region $\# 71$. Furthermore, visual inspection of the residuals clearly hints to the presence of more than 1 spectral component. 
We then fit the composite spectrum with a 2T model leaving 
normalization, temperature and Fe abundance as free parameters for both components. This results in a highly significant improvement in the fit 
($\chi^2$ of $652.5$ vs $673$ dof) and in the disappearance of the structures in the residuals found in the 1T fit. We have performed the same 2T 
fitting also on the spectra extracted from regions $\# 2$ and $\# 71$, finding that in these cases the inclusion of a second  component provides improvements that are 
significantly smaller than those found for the composite region (see Table \ref{tab:mixing}). These improvements, while formally statistically significant, as indicated by the 
results of the F-test (see Table \ref{tab:mixing}), can be ascribed to the known EPIC calibration mismatch between soft and hard spectral bands \citep[e.g.][]{Nevalainen:2010} .  

We have conducted the same analysis on other pairs of spectra finding similar results.

We conclude that while mixing of different phases can lead to spectra whose 1T best fits lie on the K-Fe correlation, such composite spectra would be easily identified
with a multi temperature analysis. The lack of any evidence of multiphaseness, beyond the modest level expected from the known EPIC calibration issues, tells us
that mixing must at the most be modest in the gas located within the spiral.

\end{document}